\documentclass[prl,twocolumn,showpacs,preprintnumbers,amsmath,amssymb,superscriptaddress]{revtex4-1}

\usepackage{graphicx}% Include figure files
\usepackage{dcolumn}% Align table columns on decimal point
\usepackage{bm}% bold math
\usepackage{epsfig}
\usepackage{amsmath}
\usepackage{amssymb}
\usepackage{color}
\usepackage{mathtools}
\usepackage{bbm}
\usepackage[caption=false]{subfig}
\usepackage{placeins}
\usepackage{soul}
\usepackage{changes} % Changes 
\definechangesauthor[name={Per cusse}, color=red]{per} % Changes by red
\usepackage[colorlinks=true,citecolor=blue,linkcolor=blue,urlcolor=blue]{hyperref}
\usepackage{cancel}

\def\equationautorefname~#1\null{%
	Eq.~#1\null}
\def\figureautorefname~#1\null{%
	Fig.~#1\null}

\begin{document}

\title{Quantum-Squeezing-Induced Point-Gap Topology and Skin Effect}
\author{Liang-Liang Wan}
\affiliation{School of Physics and Institute for Quantum Science and Engineering, Huazhong University of Science and Technology, Wuhan, 430074, China}
\affiliation{Wuhan institute of quantum technology, Wuhan, 430074, China}
\author{Xin-You L\"{u}}
\email{xinyoulu@hust.edu.cn}
\affiliation{School of Physics and Institute for Quantum Science and Engineering, Huazhong University of Science and Technology, Wuhan, 430074, China}
\affiliation{Wuhan institute of quantum technology, Wuhan, 430074, China}
\begin{abstract}
We theoretically predict the squeezing-induced point-gap topology together with a {\it symmetry-protected $\mathbb{Z}_2$ skin effect} in a one-dimensional (1D) quadratic-bosonic system (QBS). Protected by a time-reversal symmetry, such a topology is associated with a novel $\mathbb{Z}_2$ invariant (similar to quantum spin-Hall insulators), which is fully capable of characterizing the occurrence of $\mathbb{Z}_2$ skin effect. Focusing on zero energy, the parameter regime of this skin effect in the phase diagram just corresponds to a {\it real- and point-gap coexisting topological phase}. Moreover, this phase associated with the symmetry-protected $\mathbb{Z}_2$ skin effect is experimentally observable by detecting the steady-state power spectral density. Our work is of fundamental interest in enriching non-Bloch topological physics by introducing quantum squeezing, and has potential applications for the engineering of symmetry-protected sensors based on the $\mathbb{Z}_2$ skin effect. 
\end{abstract}
\maketitle

The concept of topological phases of matter has radiated from the condensed-matter physics to several fields including photonics\,\citep{Ozawa2019RevModPhys}, magnetoplasmon\,\citep{Jin2016NatComm}, mechanics\,\cite{Lisa2015PNAS, Yang2015PhysRevLett, Huber2016NatPhys, Tuo2019PRB}, cold atoms\,\citep{Zhang2018AdvPhys, Cooper2019RevModPhys}, metasurface\,\citep{Phan2019LSA, Gao2020PRX, Song2021Science}, etc. In particular, growing efforts are paid to search for distinctive topological phenomena in non-Hermitian systems\,\citep{Rudner2009PRL, Esaki2011PRB, Lee2016PRL, Leykam2017PRL, Xu2017PRL, Shen2018PhysRevLett, Yao2018PhysRevLett0, Gong2018PhysRevX, Kunst2018PRL, Zhou2019PRB, Porras2019PRL, Kawabata2019PRX, Lee2019PRB, Yokomizo2019PRL, Okuma2019PRL, Song2019PRL, Yang2020PRL, Borgnia2020PRL, Okuma2020PRL, Zhang2020PRL, Zhang2022NatComm, Longhi2022PRL, Zhu2022PRL, Zhao2019Science, Helbig2020NP, Xiao2020NatPhys, Ananya2020PNAS, Hofmann2020PRR, Weidemann2020Science, Ozturk2021Science, Wang2021Science, Liang2022PRL}. The most intriguing is the non-Hermitian skin effect\,\citep{Lee2016PRL, Yao2018PhysRevLett0}, which refers to the localization of bulk states at boundaries. Accompanied with the breakdown of the bulk-boundary correspondence, it stems from the point-gap topology where the complex-valued spectrum enclosing an energy point has a nonvanishing winding number\,\citep{Gong2018PhysRevX, Kawabata2019PRX, Porras2019PRL, Lee2019PRB, Okuma2020PRL, Zhang2020PRL}.

Squeezing of bosonic fields\,\cite{Braunstein2005RMP}, as a useful technique of quantum engineering, could not only exponentially enhance light-matter interactions\,\citep{Lu2015PRL, Qin2018PRL, Leroux2018PRL, Ge2019PRL, Zhao2019SCPMA, Zhu2020PRL, Chen2021PRL, Qin2021PRL}, but also induce instability of edge state in QBSs\,\citep{Barnett2013PRA, Galilo2015PRL, Engelhardt2016PRL, Peano2016PRX, Malz2019NatComm}. In the sense that the instability arises from the complex-valued spectrum given by a non-Hermitian matrix, the QBS is also of interest in the framework of non-Hermitian physics\,\citep{Ashida2020AP, Bergholtz2021RMP}. The topological classification for the generic QBS is established based on the Bernard-LeClair 38-fold symmetry classes\,\citep{Bernard2002}, and it predicts the topological triviality of 1D QBS in terms of zero energy\,\citep{Kawabata2019PRX}. However, the bosonic Kitaev chain exhibits an end-to-end amplification and has the analogue of Majorana zero modes\,\citep{McDonald2018PRX, Wanjura2020NatComm, Flynn2020NJP, Flynn2021PRL, Wang2022PRB, Alvaro2023arXiv}, which should be an effect of point-gap topology. Such a contradiction implies that the topological nature of QBSs still remains unclear, and solving this contradiction is fundamentally interesting in exploring the exotic topological phenomena (e.g., skin effect). 

Here, we investigate the topological origin of a 1D QBS in the thermodynamic-instability regime. By introducing an unconventional time-reversal symmetry, we discover that the squeezing can induce the appearance of point-gap topology together with a symmetry-protected $\mathbb{Z}_{2}$ skin effect in the QBS. The mechanism relies on additional symmetry enriching the topology of system. In contrast to the imaginary gauge transformation in non-Hermitian systems\,\citep{Hatano1996PRL, Hatano1997PRB, Yao2018PhysRevLett0, Lee2019PRB, Okuma2019PRL}, this skin effect corresponds to a real squeezing transformation, and it is extremely sensitive to the local perturbation that breaks the time-reversal symmetry of system. By increasing the squeezing until the point gap is open at zero energy, we also find the survival of a pair of zero modes in the open boundary condition (OBC) even if the real gap closes in the periodic boundary condition (PBC). This indicates an anomalous bulk-boundary correspondence and the appearance of a real- and point-gap coexisting topological phase. Meanwhile, the $\mathbb{Z}_{2}$ skin effect, appearing in this coexisting phase, inhibits another pair of zero modes. 

Compared with the previous works focusing on the transport amplification\,\citep{McDonald2018PRX, Wanjura2020NatComm, Wang2022PRB}, Majorana bosonic analogues together with the topological metastability\,\citep{Flynn2020NJP, Flynn2021PRL}, and non-Bloch wave behaviors\,\citep{Yokomizo2021PRB}, here we introduce an unconventional time-reversal symmetry to the QBS, and uncover the symmetry-enriched topological classification. Remarkably, we also find the real- and point-gap coexisting topological phase, and it can be identified by the steady-state power spectral density. Our work builds the connection between point-gap topology together with skin effect and quantum squeezing. It opens up a door for exploring the crossover between topological physics and quantum engineering, and offers potential applications in designing new types of  topological protected devices.

\begin{figure}
\begin{centering}
\includegraphics[width=8.5cm]{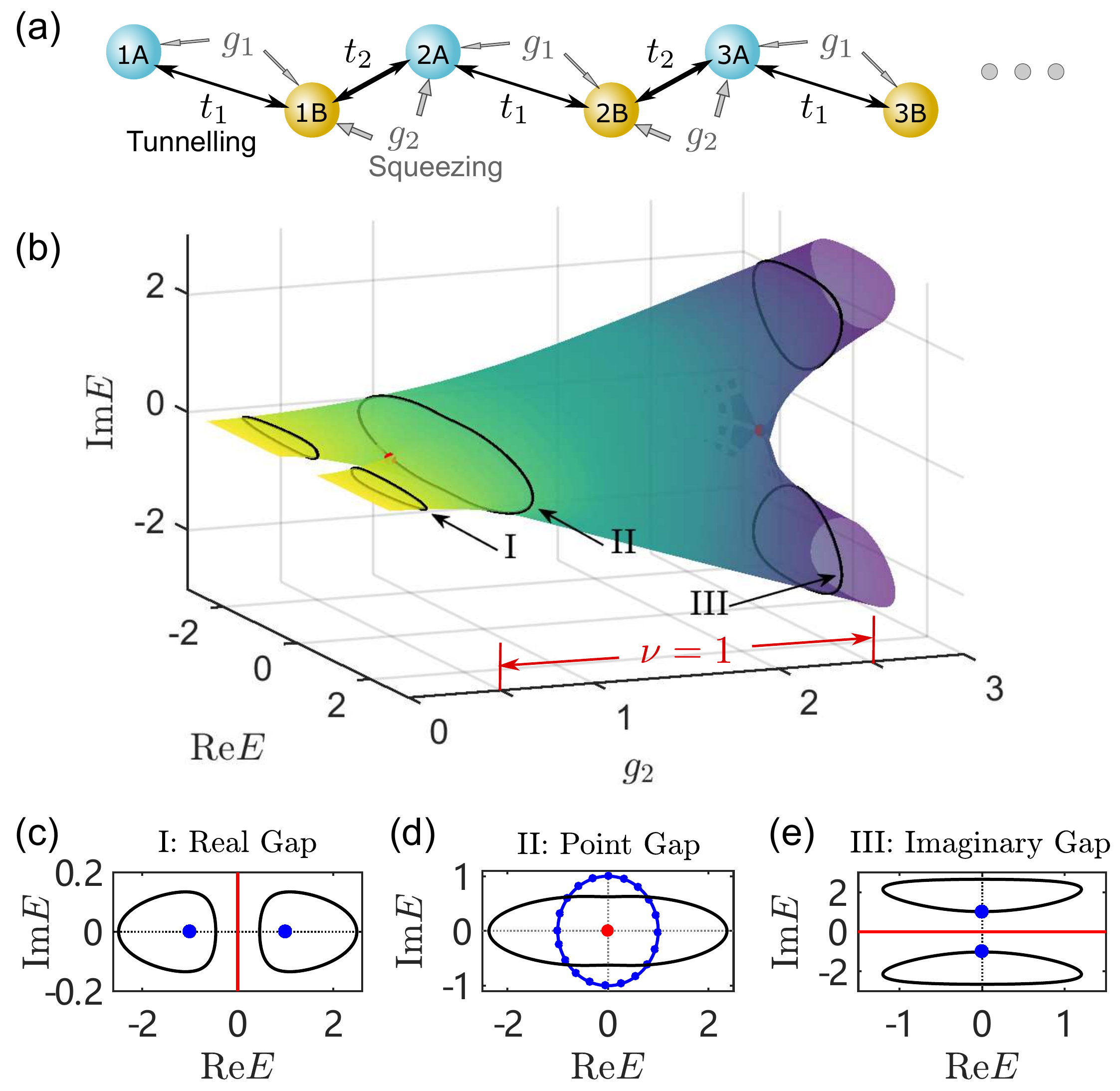}
\par\end{centering}
\caption{(a) Schematic of the squeezed SSH model consisting of the A and B sublattices in the presence of two-mode squeezing. The hopping and squeezing strengths between the adjacent sites are denoted by $t_{1}$, $t_{2}$ and $g_{1}$, $g_{2}$, respectively. (b) Spectrum in the complex plane as varying the intercell squeezing strength $g_{2}$. Here $t_{2}=3t_{1}/2$ and $g_{1}=0$. The red dots at $g_{2}=0.5t_{1},2.5t_{1}$ represent the critical points for closing or opening the point gap at $E=0$, and $\nu=1$ corresponds to $g_{2}\in\left(0.5,2.5\right)t_{1}$. (c-e) Spectra (black curves) for I, II and III in (b) can be continuously deformed to (c) $\pm1$ (blue dots), (d) unit circle (blue circle) and (e) $\pm i$ (blue dots), respectively, while preserving the associated gaps (red). \label{fig1}}
\end{figure}

\emph{Squeezing-induced point-gap topology.}---Let us consider a 1D QBS subject to the lattice-translational symmetry with Hamiltonian $\hat{H}=\frac{1}{2}\sum_k\hat{\Phi}_k^{\dagger}H(k)\hat{\Phi}_k$. Here $H(k)$ is the first-quantized Hamiltonian of the QBS in the crystal-momentum space, and $\hat{\Phi}_k=(\hat{a}_{k1},\ldots,\hat{a}_{kN},\hat{a}_{-k1}^{\dagger},\ldots,\hat{a}_{-kN}^{\dagger})^{T}$ is the Nambu spinor in terms of $2N$ bosonic annihilation and creation operators with $k$ and $-k$, respectively. The spinor obeys $[\hat{\Phi}_{ki},\hat{\Phi}_{k^\prime j}^{\dagger}]=\delta_{kk^\prime}(\tau^{3})_{ij}$ with $\tau^{3}$ being the indefinite metric\,\citep{Rossignoli2005PhysRevA,Gohberg2005}. Here, $\tau^{i}=\sigma^{i}\otimes I_{N}$ with the Pauli matrices $\sigma^{i}$ ($i=1,2,3$). The system dynamics is described by $\partial_t \hat{\Phi}_k\left(t\right)=-iH_{\tau}(k)\hat{\Phi}_k\left(t\right)$ with $H_{\tau}(k)=\tau^{3}H(k)$ being non-Hermitian.  The dynamical matrix $H_{\tau}(k)$ inherently respects the particle-hole symmetry ${\cal C}H_{\tau}^{*}(-k){\cal C}^{-1}=-H_{\tau}(k)$ with ${\cal C}=\tau^{1}$ being the ``charge conjugation''\,\citep{Bardyn2016, Peano2016NatCommun, Lieu2018PRB} and the pseudo-Hermiticity $\eta H_{\tau}^{\dagger}(k)\eta^{-1}=H_{\tau}(k)$ with $\eta=\tau^{3}$\,\citep{Mostafazadeh2001JMP}. 

In the thermodynamic-instability regime, the squeezing may induce a complex-valued spectrum formed by loops in the PBC and open curves in the OBC\,\citep{McDonald2018PRX, Wanjura2020NatComm, Okuma2022PRB}. This scenario is a reminiscence of the point-gap topology in non-Hermitian systems\,\citep{Gong2018PhysRevX, Kawabata2019PRX, Porras2019PRL}. In terms of zero energy, we construct the Hermitian matrix
\begin{equation}
\tilde{H}_{\tau}\left(k\right)=\left(\begin{array}{cc}
0 & H_{\tau}\left(k\right)\\
H_{\tau}^{\dagger}\left(k\right) & 0
\end{array}\right),
\label{eq:Hermitization}
\end{equation}
which respects the chiral symmetry $\Gamma\tilde{H}_{\tau}\Gamma^{-1}=-\tilde{H}_{\tau}$ with $\Gamma=I_{2N}\oplus-I_{2N}$. This symmetry leads to the winding number $W\in\mathbb{Z}$ given by
\begin{equation}
W=\int_{{\rm BZ}}\frac{dk}{2\pi i}\frac{\partial}{\partial k}\ln\det H_{\tau}\left(k\right).\label{eq:Winding_Spec}
\end{equation}
Equation (\ref{eq:Winding_Spec}) is always trivial due to the pseudo-Hermiticity. However, in general, the symmetry class together with the topological classification for the QBS would be altered once some additional symmetries are introduced. Hence, the presence of additional symmetries can enrich the topological phase of the QBS\,\cite{Supp}. 

For illustrations, we study the squeezed Su-Schrieffer-Heeger (SSH) model shown in Fig.\,\ref{fig1}(a). The system Hamiltonian is 
\begin{equation}
\begin{split}\hat{H}_{{\rm SSH}} =&\sum_{j\in\mathbb{Z}}\left(t_{1}\hat{a}_{j,A}^{\dagger}\hat{a}_{j,B}+t_{2}\hat{a}_{j+1,A}^{\dagger}\hat{a}_{j,B}\right.\\
 & \left.+g_{1}\hat{a}_{j,A}\hat{a}_{j,B}+g_{2}\hat{a}_{j+1,A}\hat{a}_{j,B}+{\rm H.c.}\right),
\end{split}
\label{eq:SSH_Sqz_Ham}
\end{equation}
where $t_{1},t_{2}>0$ are the hopping strengths between the nearest-neighbor sites, and $g_{1},g_{2}\in\mathbb{R}$ are the strengths of the intracell and intercell squeezing, respectively. This model can be implemented in many platforms like quantum superconducting circuits\,\citep{Abdo2013PRB, Fitzpatrick2017PRX, Frattini2017APL, Krantz2019APL, Wang2020PRX} and photonic crystals with optomechanical interaction\citep{Brooks2012Nature, safavi-naeini2013Nature, Pino2022Nature}. In particular, the crucial bosonic squeezing can be implemented via the three-wave mixing process introduced by the Josephson ring modulator or superconducting nonlinear asymmetric inductive element device\,\citep{Supp}.

The Bloch spectrum with a twofold degeneracy is   $E_{\pm}^{2}(k)=\Delta^{2}+2(t_{1}t_{2}-g_{1}g_{2})\cos k\pm2 i(t_{1}g_{2}-t_{2}g_{1})\sin k$ with $\Delta=\sqrt{t_{1}^{2}+t_{2}^{2}-g_{1}^{2}-g_{2}^{2}}$.  Figures\,\ref{fig1}(b-e) show that the spectrum experiences three processes in the complex plane as increasing $g_{2}$. First, two isolated loops are located at the real axis (I), and subsequently a curve encloses zero energy (II). Finally, two isolated loops move to the imaginary axis (III). Those processes have the real (${\rm Re}E=0$), point ($E=0$) and imaginary (${\rm Im}E=0$) gaps, respectively. This hints the appearance of nontrivial point-gap topology at zero energy in regime II induced by squeezing. 

Specifically, the winding number (\ref{eq:Winding_Spec}) for our system is trivial, when the Bogoliubov bands enclose zero energy shown in Figs.\,\ref{fig1}(b,d). However, the system also respects a sublattice symmetry ${\cal S}H_{\tau{\rm SSH}}\left(k\right){\cal S}^{-1}=-H_{\tau{\rm SSH}}\left(k\right)$ with ${\cal S}=\sigma^{3}$ being the sublattice and $H_{\tau{\rm SSH}}$ being the dynamical matrix. The combination of the particle-hole symmetry, pseudo-Hermiticity and sublattice symmetry yields an unconventional time-reversal symmetry\,\citep{Esaki2011PRB, Sato2012PTP, Lieu2020PRL}
\begin{equation}
\begin{split}
	{\cal T}H_{\tau{\rm SSH}}^{T}(-k){\cal T}^{-1}=H_{\tau{\rm SSH}}\left(k\right), 
	& \ \ \ {\cal T}{\cal T}^{*}=-I,
\end{split}
\label{eq:TRS}
\end{equation}
with ${\cal T}=i\tau^{2}\sigma^{3}$. In terms of zero energy, this symmetry supports a $\mathbb{Z}_{2}$ invariant $\nu\in\left\{ 0,1\right\} $, defined by\,\citep{Kawabata2019PRX, Supp} 
\begin{equation}
\left(-1\right)^{\nu}={\rm sgn}\left[\frac{{\rm Pf}\left(H_{\tau{\rm SSH}}\left(0\right){\cal T}\right)}{{\rm Pf}\left(H_{\tau{\rm SSH}}\left(\pi\right){\cal T}\right)}\right],\label{eq:Z2_ind}
\end{equation}
where ${\rm Pf}\left(O\right)$ denotes the Pfaffian for any skew-symmetric matrix $O$ ($O^{T}=-O$). This $\mathbb{Z}_{2}$ invariant gives the critical points at $|t_{1}\pm t_{2}|=|g_{1}\pm g_{2}|$, i.e., the red dots in Fig.\,\ref{fig1}(b), which shows a squeezing-induced nontrivial point-gap topology in regime II. Moreover, in the regimes I and III, the point-gap topology of system can also be nontrivial, if the reference energy $E$ is not zero and is placed in the closed loop\,\citep{Supp}. 

\begin{figure}
\begin{centering}
\includegraphics[width=8.5cm]{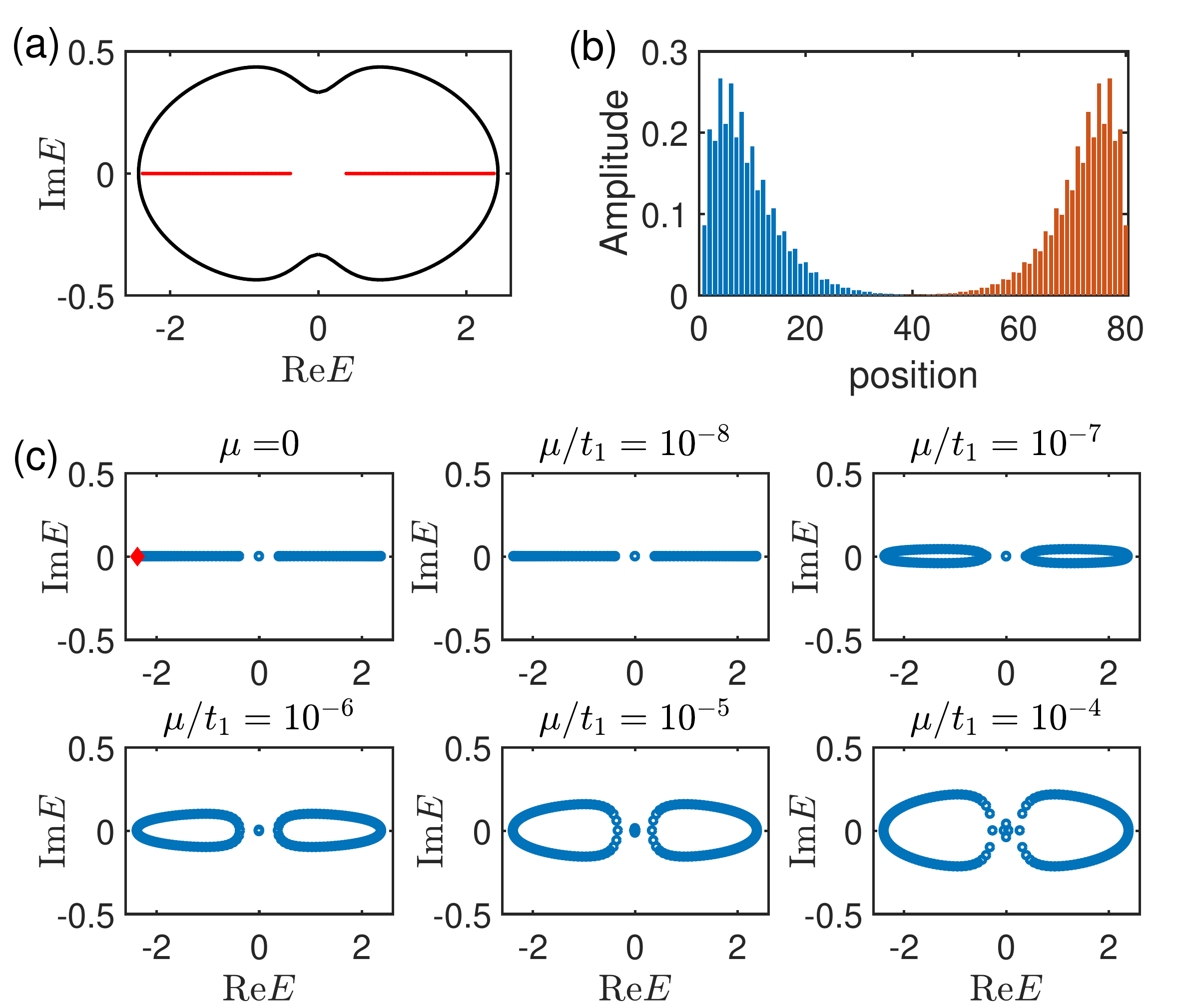}
\par\end{centering}
\caption{(a) Spectrum (black) of the squeezed SSH model under the PBC and the corresponding continuum bands (red) after the mapping (\ref{eq:Sqz_Action}). (b) Amplitudes of the Kramers pair with the lowest energy (blue and red bars) for both the particles and holes in the OBC. The localization of the two degenerate states manifests the $\mathbb{Z}_{2}$ skin effect. (c) Spectra of the perturbed model in the OBC with varying the chemical potential $\mu=(0,10^{-8},10^{-7},10^{-6},10^{-5},10^{-4})t_{1}$. The red diamond mark denotes the Kramers pair in (b). Parameters: $t_{2}=1.5t_{1}$, $g_{1}=0$, $g_{2}=0.6t_{1}$ and $L=40$. 
\label{fig2}}
\end{figure}

\emph{Symmetry-protected $\mathbb{Z}_{2}$ skin effect.}---In the presence of point-gap topology, the spectrum of Hamiltonian (\ref{eq:SSH_Sqz_Ham}) dramatically changes from a closed curve [black loop in Fig.\,\ref{fig2}(a)] to the discrete points that form open lines [see the first panel of Fig.\,\ref{fig2}(c)] under the OBC. Consequently, as shown in Fig.\,\ref{fig2}(b), the Kramers pair guaranteed by the time-reversal symmetry (\ref{eq:TRS}) are localized at both ends, which shows the appearance of the symmetry-protected $\mathbb{Z}_{2}$ skin effect\,\citep{Okuma2020PRL, Okuma2020PRB, Kawabata2020PRB}. 

In general, the non-Hermitian skin effect corresponds to an imaginary gauge transformation\,\citep{Hatano1996PRL,Hatano1997PRB,Yao2018PhysRevLett0,Lee2019PRB,Okuma2019PRL}. However, here the $\mathbb{Z}_{2}$ skin effect corresponds to a real squeezing transformation with operator $\hat{S}$\,\citep{Supp}. Specifically, under the parameter condition of $g_{1}=0$ and $t_{2}>\left|g_{2}\right|$, we perform a squeezing transformation to the ``particles'' $\hat{a}_{j\sigma}$ and ``holes'' $\hat{a}_{j\sigma}^{\dagger}$  with $\sigma=A,B$ such that 
\begin{equation}
	\left(\begin{array}{c}
		\hat{a}_{j,A/B}\\
		\hat{a}_{j,A/B}^{\dagger}
	\end{array}\right)=(e^{\pm r\tau^{1}})^{j}\left(\begin{array}{c}
		\hat{\alpha}_{j,A/B}\\
		\hat{\alpha}_{j,A/B}^{\dagger}
	\end{array}\right).
	\label{eq:Sqz_Action}
\end{equation}
Here the squeezing parameter $r$ satisfies $\tanh r=-g_{2}/t_{2}$. The squeezing transformation (\ref{eq:Sqz_Action}) inherently belongs to SU(1,1)\,\citep{Pseudounitarity}, and the particles and holes $(\hat{\alpha}_{j\sigma},\hat{\alpha}_{j\sigma}^{\dagger})$ in the new quasi-particle basis preserve  $[\hat{\alpha}_{j\sigma}, \hat{\alpha}_{j^{\prime}\sigma^{\prime}}^{\dagger}] =\delta_{jj^{\prime}}\delta_{\sigma\sigma^{\prime}}$. Using this transformation (\ref{eq:Sqz_Action}), the Hamiltonian (\ref{eq:SSH_Sqz_Ham}) is mapped to the conventional SSH model with Hamiltonian $\hat{H}_{{\rm SSH}}=\sum_{j=1}^{L}t_{1}\hat{\alpha}_{jA}^{\dagger}\hat{\alpha}_{jB}+\tilde{t}_{2}\hat{\alpha}_{j+1A}^{\dagger}\hat{\alpha}_{jB}+{\rm H.c.}$, where $\tilde{t}_{2}=\sqrt{t_{2}^{2}-g_{2}^{2}}$ and $L$ is the number of the total unit cells. As shown in Fig.\,\ref{fig2}(a), the spectrum of $\hat{H}_{{\rm SSH}}$ becomes two open (red) lines in the continuum limit $L\rightarrow\infty$ corresponding to the PBC\,\citep{Continuum}, which indicates the disappearance of skin effect in the squeezed-state representation. This demonstrates that the obtained $\mathbb{Z}_{2}$ skin effect originally comes from the intercell squeezing $\sum_{j}g_2(\hat{a}_{jB}\hat{a}_{j+1A}+{\rm H.c.})$ in the QBS. 

Physically, such squeezing interaction describes a nondegenerate parametric amplification process, and gives rise to the entanglement between two bosonic modes in the adjacent unit cells. Then, the introduced intercell parametric amplification in the 1D lattice induces intrinsically the non-Hermicity of system, which ultimately leads to the appearance of the point-gap topology together with symmetry-protected $\mathbb{Z}_2$ skin effect\,\citep{Supp}.

This $\mathbb{Z}_{2}$ skin effect is extremely sensitive against local symmetry-breaking perturbations\,\citep{Okuma2019PRL, Kawabata2020PRB, Supp}. To show this, we introduce an onsite perturbation $\hat{H}_{{\rm onsite}} = \mu\sum_{j\sigma} \hat{a}_{j\sigma}^{\dagger} \hat{a}_{j\sigma}$ to the system, which breaks the time-reversal symmetry (\ref{eq:TRS}). Applying the squeezing transformation (\ref{eq:Sqz_Action}) to the perturbation, we obtain 
\begin{equation}
\begin{split}\!\!\!\hat{H}_{{\rm onsite}} =& \mu\sum_{j=1}^{L}\left[\cosh\left(2rj\right)\left(\hat{\alpha}_{jA}^{\dagger}\hat{\alpha}_{jA}+\hat{\alpha}_{jB}^{\dagger}\hat{\alpha}_{jB}\right)\right.
\\
 & \left.+\frac{\sinh\left(2rj\right)}{2}\left(\hat{\alpha}_{jA}^{\dagger}\hat{\alpha}_{jA}^{\dagger}-\hat{\alpha}_{jB}^{\dagger}\hat{\alpha}_{jB}^{\dagger}+{\rm H.c.}\right)\right].\!\!\!
\end{split}
\label{eq:Perturb}
\end{equation}
The impact of Eq.\,(\ref{eq:Perturb}) on the unperturbed Hamiltonian is qualitatively determined by the scaling\,\citep{Okuma2019PRL, Li2020NatComm, Yokomizo2021PRB} 
\begin{equation}
\mu/t_{1}\sim e^{-\left|r\right|L}.
\label{eq:Perturb_Bound}
\end{equation}
It implies that the presence of an infinitesimal perturbation also can change the physics of system in the continuum limit. Such an instability arises from the breakdown of the time-reversal symmetry. More precisely, the anomalous squeezing in (\ref{eq:Perturb}) dramatically alters the spectrum by coupling the Kramers pairs localized at the opposite ends of the chain [see Fig.~\ref{fig2}(b)]. As shown in Fig.\,\ref{fig2}(c), the instability of spectrum occurring at $\mu/t_{1}\sim 10^{-8}$ confirms our analysis. 
\begin{figure}
\begin{centering}
\includegraphics[width=7.5cm]{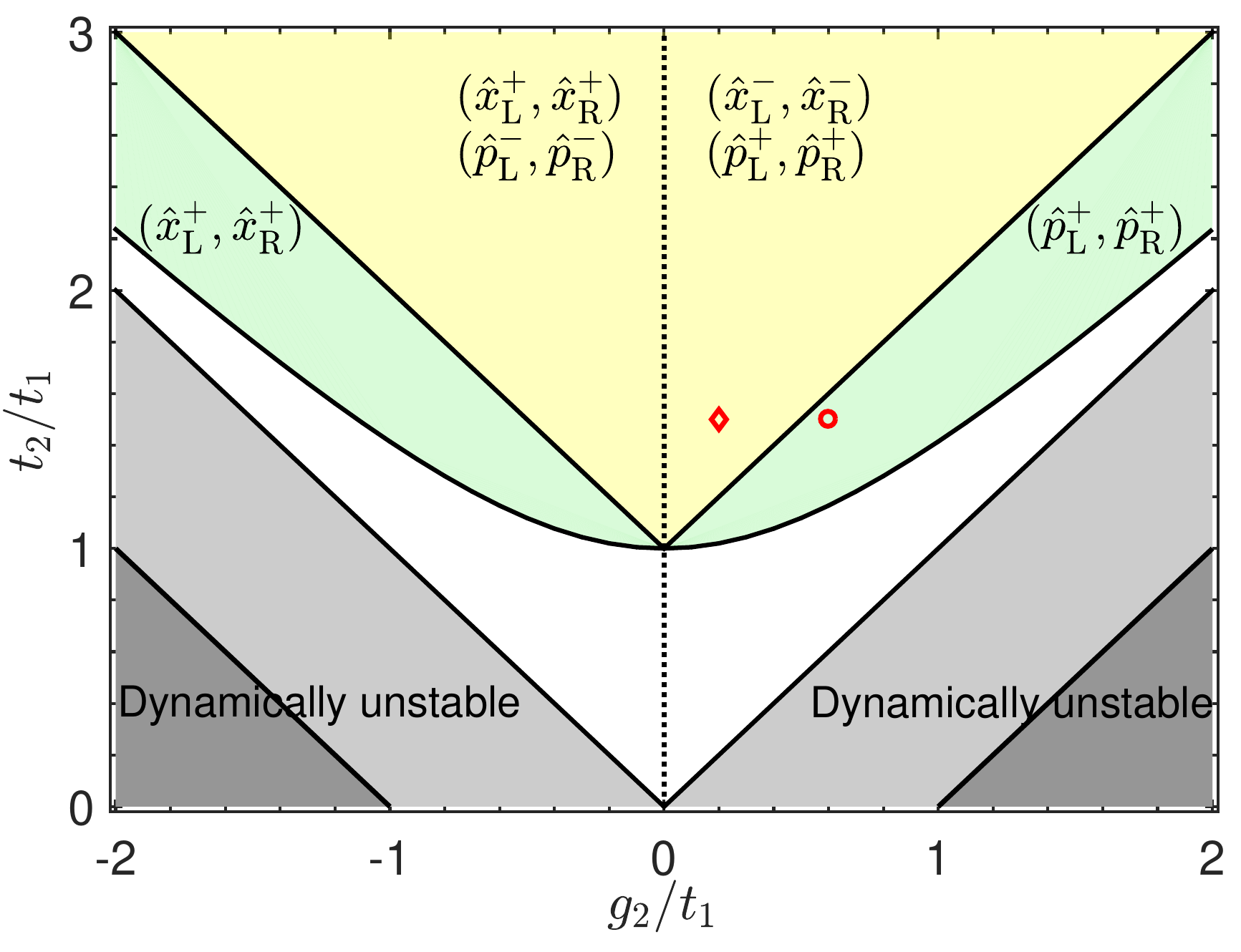}
\par\end{centering}
\caption{Phase diagram of the 1D QBS for $g_{1}=0$. In the yellow area, the real gap opens in the PBC, and two pairs of zero modes $(\hat{x}_{\rm L}^{\pm},\hat{x}_{\rm R}^{\pm})$ and $(\hat{p}_{\rm L}^{\mp},\hat{p}_{\rm R}^{\mp})$ appear in the OBC. The green area indicates a real- and point-gap coexisting topological phase with a zero-mode pair. The dark gray areas correspond to the imaginary-gap topological phase. The system has no zero mode in the white area, and is dynamically unstable in the light and dark gray areas under the OBC. \label{fig3}}
\end{figure}

\emph{Real- and point-gap coexisting topological phase.}---The parameter regime of $\mathbb{Z}_{2}$ skin effect actually corresponds to a real- and point-gap coexisting topological phase due to the interplay between the squeezing and particle-exchange coupling. Such a phase is unconventional since the real gap is closed in the PBC while the zero mode survives in the OBC, which indicates an anomalous bulk-boundary correspondence. Meanwhile, the point-gap topology is also nontrivial. To show this, in Fig.\,\ref{fig3}, we plot the phase diagram for the real-gap topology by calculating the winding number in the PBC and the zero modes in the OBC. 

Firstly, the real gap ${\rm Re}E=0$ opens in the PBC if $|g_{2}|<|t_{2}-t_{1}|$ holds, as shown in Fig.\,\ref{fig1}(b). Because of the sublattice symmetry ${\cal S}$, the real-gap topology can be characterized by the winding number $W^{({\rm real})}=(1/2\pi i)\int_{{\rm BZ}}q^{-1}dq$ with $q=t_{1}+t_{2}e^{ik}$\,\citep{Supp}. This winding number is nontrivial for $t_{1}+|g_{2}|<t_{2}$, corresponding to the yellow area of Fig.\,\ref{fig3}. The bulk-boundary correspondence ensures the emergence of zero modes in the bulk gap. In the representation of the canonical coordinates and momenta  $\hat{x}_{j\sigma}=(\hat{a}_{j\sigma}+\hat{a}_{j\sigma}^{\dagger})/\sqrt{2}$ and $\hat{p}_{j\sigma}=(\hat{a}_{j\sigma}-\hat{a}_{j\sigma}^{\dagger})/\sqrt{2}i$, two pairs of zero modes $\hat{x}_{\rm L}^{-s}=\sum_{j=1}^{L}\delta_{-s}^{j-1}\hat{x}_{jA}$, $\hat{x}_{\rm R}^{-s}=\sum_{j=1}^{L}\delta_{-s}^{L-j}\hat{x}_{jB}$ and $\hat{p}_{\rm L}^{s}=\sum_{j=1}^{L}\delta_{s}^{j-1}\hat{p}_{jA}$, $\hat{p}_{\rm R}^{s}=\sum_{j=1}^{L}\delta_{s}^{L-j}\hat{p}_{jB}$ with $\delta_{\pm s}=-t_{1}/(t_{2}\pm s|g_{2}|)$ and $s={\rm sgn}(g_{2})=\pm$ ($|\delta_{\pm}|<1$) appear in the OBC\,\citep{Supp}. Here the subscripts L and R denote the left and right edges of the 1D QBS, respectively. $[\hat{x}^{-s}_{\rm L},\hat{p}^{s}_{\rm L}]=[\hat{x}^{-s}_{\rm R},\hat{p}^{s}_{\rm R}]= i(1-\delta^{2L})/(1-\delta^2)$ ($\delta=-t_1/\tilde{t}_2$) implies that $\hat{x}^{-s}_{\rm L/R}$ and $\hat{p}^{s}_{\rm L/R}$ are canonically conjugate with each other.
 
As increasing $g_2$, the real gap closes at $t_{1}+|g_{2}|=t_{2}$, while a pair of zero mode $(\hat{x}_{\rm L}^{+},\hat{x}_{\rm R}^{+})$ or $(\hat{p}_{\rm L}^{+},\hat{p}_{\rm R}^{+})$ can survive. This means that the conventional bulk-boundary correspondence based on $W^{({\rm real})}$ is no longer valid. To reconstruct it, we impose the continuum limit to the mapped Hamiltonian $\hat{H}_{{\rm SSH}}$, and find that the real gap preserves in the region $|g_{2}|<t_{2}$ under the PBC. Furthermore, the reconstructed winding number $\tilde{W}^{({\rm real})}$\,\citep{Supp} indicates the new nontrivial phase (i.e., $\sqrt{t_{1}^{2}+g_{2}^{2}}<t_{2}$), corresponding to the yellow and green areas of Fig.\,\ref{fig3}. 

In terms of $E=0$, the defined $\nu$ is nontrivial in the green area, which indicates a real- and point-gap coexisting topological phase. Correspondingly, the symmetry-protected $\mathbb{Z}_2$ skin effect appears and greatly inhibits the occurrence of a pair of zero modes, either $(\hat{x}_{\rm L}^{-},\hat{x}_{\rm R}^{-})$ for $g_2>0$ or $(\hat{p}_{\rm L}^{-},\hat{p}_{\rm R}^{-})$ for $g_2<0$. This inhibition originates from the localization competition between the skin effect and zero modes of the conventional SSH model\,\cite{Supp}. Meanwhile, another pair of zero modes $(\hat{p}_{\rm L}^{+},\hat{p}_{\rm R}^{+})$ or $(\hat{x}_{\rm L}^{+},\hat{x}_{\rm R}^{+})$ survive, and they are extremely sensitive to the local perturbation (\ref{eq:Perturb}). The scaling of $\mu$ can be heuristically estimated by $\mu/t_1\sim \xi^{-L}$ with $\xi=e^{\left|r\right|}|\delta|^{1/2}$. Figure\,\ref{fig2}(c) shows that the zero modes $(\hat{p}_{\rm L}^{+},\hat{p}_{\rm R}^{+})$ almost disappear at $\mu/t_{1}\sim 3\times10^{-5}$, which is consistent with this critical scaling. As continuously increasing $g_2$, the imaginary gap is open and the associated topology becomes nontrivial in regime III of Figs.\,\ref{fig1}(b,e), corresponding to the dark gray areas of Fig.\,\ref{fig3}. Moreover, the phase diagram can be enriched further when the intracell squeezing is introduced, i.e., $g_{1}\neq0$\,\citep{Supp}.

\begin{figure}
\begin{centering}
\includegraphics[width=8.5cm]{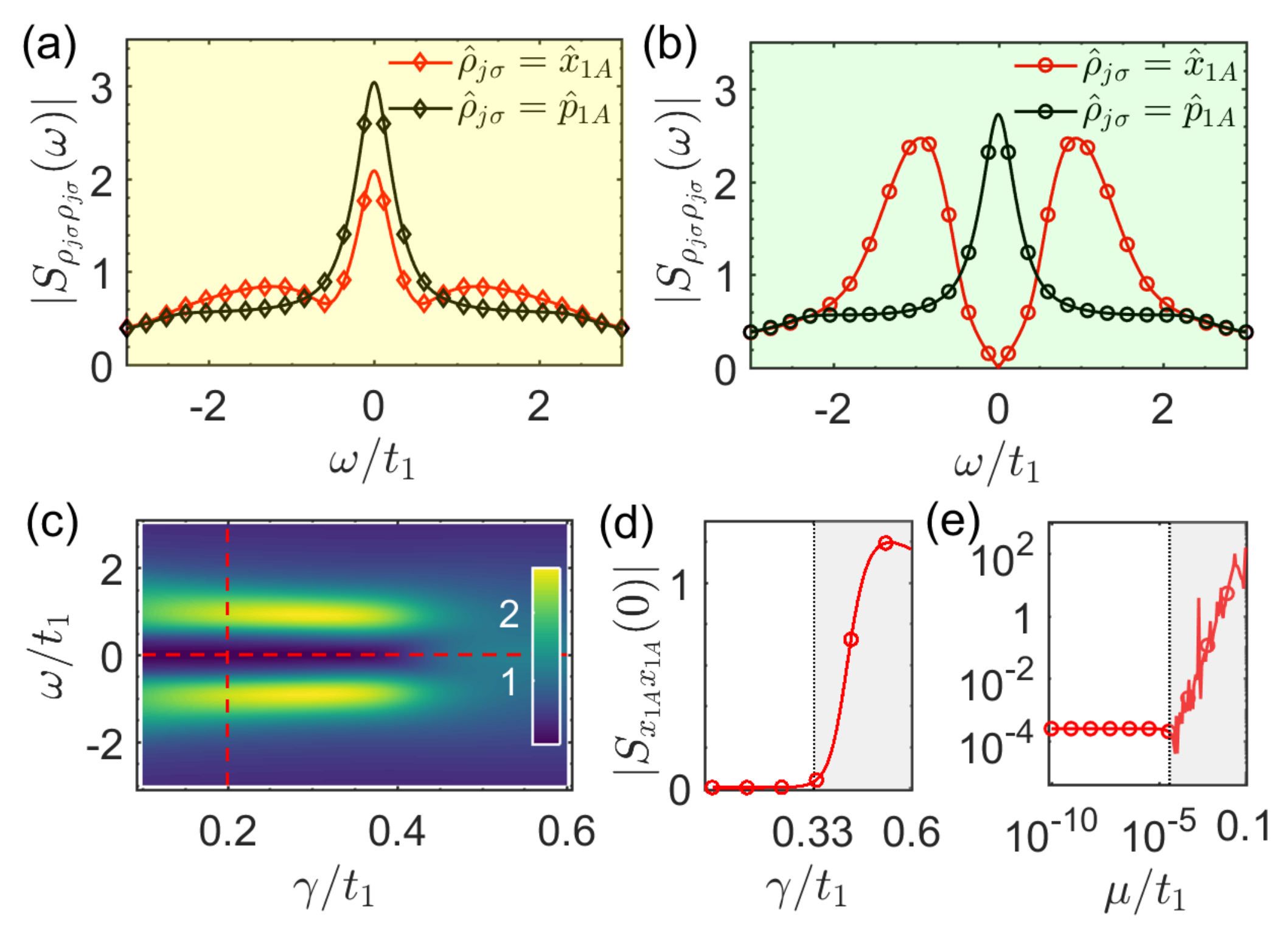}
\par\end{centering}
\caption{(a,b) Power spectral density $|S_{\rho_{j\sigma}\rho_{j\sigma}}(\omega)|$ with $\hat{\rho}_{j\sigma}=\hat{x}_{1A}, \hat{p}_{1A}$ versus $\omega$ when $\gamma=0.2t_{1}$, (a) $g_{2}=0.2t_{1}$ and (b) $g_{2}=0.6t_{1}$, corresponding to the diamond and circle in Fig.\,\ref{fig3}. (c) $|S_{x_{1A}x_{1A}}(\omega)|$ as varying $\gamma$ and $\omega$. The red dashed lines denote the cases in (b) and (d), respectively. (d) $|S_{x_{1A}x_{1A}}(0)|$ versus $\gamma$. The zero-frequency dip vanishes at the critical decay $\gamma_{c}\approx0.33t_{1}$. (e) $|S_{x_{1A}x_{1A}}(0)|$ versus $\mu$ when $\gamma=0.2t_{1}$. The topological inhibition vanishes at $\mu/t_{1}\sim 3\times 10^{-5}$. Parameters: $t_{2}=1.5t_{1}$, $g_{1}=0$, $L=40$, and (c-e) $g_{2}=0.6t_{1}$. \label{fig4}}
\end{figure}

\emph{Detection of the coexisting topological phase together with the $\mathbb{Z}_{2}$ skin effect.}--- For detection, we calculate the normalized power spectral density $S_{\rho_{j\sigma}\rho_{j\sigma}}(\omega)=\int{\rm d}\tau\langle\hat{\rho}_{j\sigma}(\tau)\hat{\rho}_{j\sigma}(0)\rangle_{{\rm ss}}e^{i\omega\tau}/\langle\hat{\rho}_{j\sigma}(0)\hat{\rho}_{j\sigma}(0)\rangle_{{\rm ss}}$ in the presence of decay with rate $\gamma$\,\citep{Supp}. Here $\langle\cdot\rangle_{{\rm ss}}$ denotes a steady-state expectation value and $\hat{\rho}_{j\sigma}=\hat{x}_{j\sigma},\hat{p}_{j\sigma}$. Normally, any zero mode corresponds to the peak of $|S_{\rho_{j\sigma}\rho_{j\sigma}}(0)|$ at edge sites. Focusing on the first site 1A, the zero modes $\hat{x}^{-s}_{\rm L}$ and $\hat{p}^{s}_{\rm L}$ correspond to the peaks of $|S_{x_{1A}x_{1A}}(0)|$ and $|S_{p_{1A}p_{1A}}(0)|$, respectively. Then double peaks at zero frequency in Fig.\,\ref{fig4}(a) indicate the real-gap topological phase (yellow area of Fig.\,\ref{fig3}), and one peak in Figs.\,\ref{fig4}(b) depicts the real- and point-gap coexisting topological phase (green area in Fig.\,\ref{fig3}). Moreover, the peaks of $|S_{x_{1A}x_{1A}}(\pm t_{1})|$ in Fig.\,\ref{fig4}(b) also manifest the skin effect, which is algebraically divergent with $L$\,\citep{Supp, Flynn2021PRL}. 

The above signature for detecting the coexisting topological phase (i.e., the zero-frequency dip) will be destroyed by the dissipation or perturbation of the system. Figures\,\ref{fig4}(c,d) show that the dip of $|S_{x_{1A}x_{1A}}(0)|$ disappears at the critical point $\gamma_{c}\equiv\sqrt{g_{2}^{2}-(t_{1}-t_{2})^{2}}$\,\citep{Supp}. Physically, the presence of dissipation moves the effective spectrum to the lower half plane, and the reference frequency $\omega$ would go out of the loop as increasing $\gamma$. Figure\,\ref{fig4}(e) demonstrates that the zero-frequency dip vanishes at the scaling $\mu/t_1\sim\xi^{-L}$, since the perturbation breaks the time-reversal symmetry. 

\emph{Conclusion.}---We have shown the squeezing-induced point-gap topology together with the $\mathbb{Z}_{2}$ skin effect in the QBS, when the time-reversal symmetry is introduced. The interplay of the bosonic squeezing and particle-exchange coupling results in the survival of zero modes in the OBC even if a real gap closes in the PBC. This exhibits an anomalous bulk-boundary correspondence. Our work enriches non-Bloch topological physics in the QBS by predicting the real- and point-gap coexisting topological phase. This may stimulate future studies of symmetry-enriched topological physics in the higher-dimensional systems. Our work also provides a perfect example of the combination of non-linearity and non-Hermiticity with topology, and it will inspire experimental activity in the field of nonlinear topological photonics\,\citep{Smirnova2020APR}. 

L.-L.W. is very thankful to Dr.\,Zixian Zhou for his fruitful discussions. This work is supported by the National Key Research and Development Program of China (Grant No.\,2021YFA1400700), the National Natural Science Foundation of China (Grants No.\,11974125, No.\,12205109, No.\,12147143).

%\bibliography{Sqz_Top}
%\bibliographystyle{apsrev4-1}

%

\setcounter{secnumdepth}{2}
\onecolumngrid
\clearpage
\setcounter{equation}{0}
\setcounter{figure}{0}
\setcounter{table}{0}
\setcounter{page}{6}
\setcounter{section}{0}

\makeatletter
\renewcommand{\theequation}{S\arabic{equation}}
\renewcommand{\thefigure}{S\arabic{figure}}
\renewcommand{\bibnumfmt}[1]{[S#1]}
\renewcommand\thesection{\Roman{section}}

\begin{center}
\Large \textbf{Supplemental Material for ``Quantum-Squeezing-Induced Point-Gap Topology and Skin Effect''}
\end{center}

\title{Supplemental Material for ``Quantum-Squeezing-Induced Point-Gap Topology and Skin Effect''}
\author{Liang-Liang Wan}
\affiliation{School of Physics and Institute for Quantum Science and Engineering, Huazhong University of Science and Technology, Wuhan, 430074, China}
\affiliation{Wuhan institute of quantum technology, Wuhan, 430074, China}
\author{Xin-You L\"{u}}
\email{xinyoulu@hust.edu.cn}
\affiliation{School of Physics and Institute for Quantum Science and Engineering, Huazhong University of Science and Technology, Wuhan, 430074, China}
\affiliation{Wuhan institute of quantum technology, Wuhan, 430074, China}
\date{\today} 
\maketitle

\begin{center}
Liang-Liang Wan$^{1,2}$, Xin-You L\"{u}$^{1,2,*}$
\end{center}

\begin{minipage}[]{16cm}
{\it
\centering ${}^{1}$School of Physics and Institute for Quantum Science and Engineering,\\
Huazhong University of Science and Technology, Wuhan, 430074, China \\
\centering $^2$Wuhan institute of quantum technology, Wuhan, 430074, China\\}
\end{minipage}

\vspace{8mm}

In this supplemental material, we provide some details to support the main text: Section\,\ref{sec:SM_Symm_enrich_Top} demonstrates concretely the point-gap topology enriched by additional symmetry in one-dimensional (1D) quadratic bosonic systems (QBSs). In Sec.\,\ref{sec:SM_SSH}, the topological invariant under the periodic boundary condition (PBC) for the squeezed Su-Schrieffer-Heeger (SSH) model is derived. Then a detailed derivation of the $\mathbb{Z}_2$ skin effect and zero modes under the open boundary condition (OBC) is provided in Sec.\,\ref{sec:SM_SkinEffect}. In Sec.\,\ref{sec:SM_NonBloch}, we also discuss the reconstructed real-gap topology and the infinitesimal instability of the $\mathbb{Z}_2$ skin effect in detail. Section\ref{sec:SM_Imag_Top} discusses the case of the imaginary-gap topology for completeness. In Sec.\,\ref{sec:SM_Power_Spec}, we give an additional discussion for detecting topology via the power spectral density. Section\,\ref{sec:SM_Proof} is devoted to a general proof on the correspondence between the squeezing transformation and $\mathbb{Z}_2$ skin effect in QBSs. After that, we provide an additional discussions on the intercell squeezing in Sec.\,\ref{sec:SM_ISqz}. Finally, the physical implementation of squeezed SSH model is discussed in Sec.\,\ref{sec:SM_Implementation}.

\section{Symmetry-enriched point-gap topology \label{sec:SM_Symm_enrich_Top}} 
In this section, we provide further details on how the additional symmetries may alter the symmetry class, and hence, the topological classification of the QBS claimed in the main text. Let us focus on the 1D QBS with the lattice-translational symmetry to study the point-gap topology. The dynamical matrix of the 1D QBS $H_{\tau}\left(k\right)$ takes the form 
\begin{equation}
	H_{\tau}\left(k\right)=\left(\begin{array}{cc}
		H_{0}\left(k\right) & Z\left(k\right)\\
		-Z^{*}\left(-k\right) & -H_{0}^{*}\left(-k\right)
	\end{array}\right),
\end{equation}
where $H_{0}\left(k\right)=H_0^\dagger(k)$ describes the particle-conserving part and $Z\left(k\right)=Z^{T}\left(-k\right)$ is the squeezing part.  The dynamical matrix $H_\tau(k)$ inherently respects the particle-hole symmetry and pseudo-Hermiticity, i.e.,
\begin{align}
	{\cal{C}}H_\tau^*(-k){\cal{C}}^{-1} =-H_\tau(k)\,\,\, {\rm and}\,\,\, & \eta H_\tau^\dagger(k)\eta^{-1} =H_\tau(k),
\end{align}
where ${\cal{C}}=\tau^1$ and $\eta=\tau^3$, respectively. For the system gapped at zero energy, the dynamical matrix satisfies $\det\left(H_{\tau}\left(k\right)\right)\neq0$. Then $H_{\tau}\left(k\right)$ can be continuously deformed to a unitary matrix in terms of zero energy \citep{Gong2018PhysRevX}, 
\[
H_{\tau,\lambda}\left(k\right)\coloneqq\left(1-\lambda\right)H_{\tau}\left(k\right)+\lambda U_{\tau}\left(k\right),\ \ \ \lambda\in\left[0,1\right]
\]
with unitary matrix $U_{\tau}\left(k\right)\coloneqq H_{\tau}\left(k\right)P^{-1}$ and $P\left(k\right)\coloneqq\sqrt{H_{\tau}\left(k\right)H_{\tau}^{\dagger}\left(k\right)}$ being positive definite. Such a dynamical matrix $H_{\tau}\left(k\right)$ is said to be homotopically equivalent to $U_{\tau}\left(k\right)$, denoted by $H_{\tau}\left(k\right)\simeq U_{\tau}\left(k\right)$. One notes that the topological classification of the unitary matrix is equivalent to an Hermitian matrix respecting a chiral symmetry\,\citep{Kitaev2009AIP}. Therefore, rather than $U_\tau(k)$, we can turn to the Hermitian matrix
\begin{equation}
	\tilde{H}_\tau\left(k\right)=\left(\begin{array}{cc}
		0 & U_{\tau}\left(k\right)\\
		U_{\tau}^{\dagger}\left(k\right) & 0
	\end{array}\right)\simeq\left(\begin{array}{cc}
		0 & H_{\tau}\left(k\right)\\
		H_{\tau}^{\dagger}\left(k\right) & 0
	\end{array}\right).\label{eq:SM_Hermitization}
\end{equation}
The Hermitian matrix respects an additional chiral symmetry, i.e., $\Gamma \tilde{H}_\tau(k)\Gamma^{-1}=-\tilde{H}_\tau(k)$ with 
\begin{equation}
	\Gamma=\left(\begin{array}{cc}
		I_{2N} & 0\\
		0 & -I_{2N}
	\end{array}\right).\label{eq:SM_Chiral_Sym}
\end{equation}
With this symmetry, one can define the winding number $W$, which is always trivial for the QBS, i.e.,
\begin{equation}
	\begin{split}W & =\frac{1}{2\pi i}\int_{{\rm BZ}}d\ln\left[\det\left(U_{\tau}\left(k\right)\right)\right]\\
		& =\frac{1}{2\pi i}\int_{{\rm BZ}}d\ln\left[\det\left(H_{\tau}\left(k\right)\right)\right]\\
		& =\frac{1}{2\pi i}\int_{{\rm BZ}}d\ln\left[\det\left(H_{\tau}^{*}\left(k\right)\right)\right]\\
		& \equiv0.
	\end{split}
	\label{eq:SM_Winding_Spec}
\end{equation}
In the third line, we have used the pseudo-Hermiticity of the QBS. In Ref.\,\citep{Kawabata2019PRX}, this QBS is classified into class C with $\eta{\cal C}=-\eta{\cal C}$ in the absence of time-reversal symmetry, and class CI (CII) with $\eta{\cal T}={\cal T}\eta$ and $\eta{\cal C}=-\eta{\cal C}$ in the presence of time-reversal symmetry ${\cal T}H_\tau^*(-k){\cal T}=H_\tau(k)$ with ${\cal T}{\cal T}^{*}=+I$ (${\cal T}{\cal T}^{*}=-I$). In these classes, the 1D QBS is predicted to be topologically trivial, which coincides with Eq.\,(\ref{eq:SM_Winding_Spec}). However, the symmetry class of the QBS can be generally altered once some additional
and physical symmetries are introduced. Thus, the topological classification of such systems can be changed and the symmetry-enriched topological phase may occur in the QBS. We provide two cases to support our argument: the presence of the additional (i) particle-hole symmetry and (ii) sublattice symmetry. 

\subsection{Additional particle-hole symmetry }
We assume that an additional particle-hole symmetry ${\cal C}^{\prime}$, i.e., 
\begin{align}
	{\cal C}^{\prime}H_{\tau}^{*}\left(-k\right){\cal C}^{\prime -1}=-H_{\tau}\left(k\right),\ \ \ & {\cal C}^{\prime}{\cal C}^{\prime*}=\pm I,
\end{align}
is introduced into the system. Combining with the inherent particle-hole symmetry ${\cal C}$, we obtain a ``parity'' symmetry 
\begin{equation}
	{\cal P}_{{\cal C}}H_{\tau}\left(k\right){\cal P}_{{\cal C}}^{-1}=H_{\tau}\left(k\right),
\end{equation}
with ${\cal P}_{{\cal C}}\coloneqq{\cal C}{\cal C}^{\prime}$. The unitary ${\cal P}_{{\cal C}}$ divides the space consisting of the eigenstates of $H_\tau(k)$ into two subspaces with the opposed parity $\pm1$,
respectively. In the diagonal form, the dynamical matrix becomes 
\begin{equation}
	H_{\tau}^{\left({\rm diagonal}\right)}\left(k\right)=\left(\begin{array}{cc}
		H_{+}\left(k\right) & 0\\
		0 & H_{-}\left(k\right)
	\end{array}\right).
\end{equation}
The diagonal blocks are related to each other and no longer have the particle-hole symmetry. As a result, one can focus on the block $H_{+}\left(k\right)$ to study the topology of the QBS without loss of generality. Note that the pseudo-Hermiticity is also trivialized in each diagonal block for physical reasons. In this case, $H_+(k)$ has no symmetry when the QBS has no time-reversal symmetry. Applying Eq.\,(\ref{eq:SM_Hermitization}) to $H_+(k)$, we find the fact that the QBS subject to a point gap should be classified into class AIII in the absence of time-reversal symmetry. Accordingly, the classifying space for the 1D QBS is changed. And the topological classification for such systems is the collection of integers ($\mathbb{Z}$). Thus, 1D QBSs can be topologically nontrivial due to the presence of ${\cal C}'$. Moreover, in the presence of time-reversal symmetry ${\cal T}{\cal T}^{*}=\pm I$, the QBS should be classified into class BDI and CII (DIII and CI) for ${\cal T}\Gamma=\Gamma{\cal T}$ (${\cal T}\Gamma=-\Gamma{\cal T}$), respectively. The associated topological classification for the 1D QBS can be seen in Ref.\,\citep{Chiu2016RevModPhys}.

For instance, we revisit the bosonic Kitaev chain in the presence of onsite single-mode squeezing, and the Hamiltonian in the $k$-space is given by\,\citep{McDonald2018PRX,Flynn2020NJP}
\begin{equation}
	H_{{\rm BK}}\left(k\right)=t\sin k\tau^{3}-\left(\mu^{\prime}+\Delta^{\prime}\cos k\right)\tau^{2},
\end{equation}
where $t$ is the hopping rate and $\Delta^{\prime}$ ($\mu^{\prime}$) is the strength of the two- (single-)mode squeezing. Apart from the inherent particle-hole symmetry ${\cal C}=\tau^{1}$, the dynamical matrix also respects another particle-hole symmetry 
\begin{equation}
	{\cal C}'H_{\tau{\rm BK}}^{*}\left(-k\right){\cal C}'^{-1}=-H_{\tau{\rm BK}}\left(k\right),
\end{equation}
where ${\cal C}'=I$. Combining them two, this model respects a parity symmetry. And in the representation of the parity ${\cal P}_{\cal C}$, the dynamical matrix is diagonal 
\begin{equation}
	\begin{split}\tilde{H}_{\tau{\rm BK}}\left(k\right) & =\left(\begin{array}{cc}
			H_{{\rm \tau BK}}^{(+)}\left(k\right) & 0\\
			0 & H_{{\rm \tau BK}}^{(-)}\left(k\right)
		\end{array}\right),\end{split}
\end{equation}
where $H_{{\rm \tau BK}}^{(\pm)}\left(k\right)=t\sin k\pm i\left(\mu^{\prime}+\Delta^{\prime}\cos k\right)$. It can be seen that the point-gap  topology of the non-Hermitian matrix
is determined by the diagonal entry of the matrix $H_{{\rm\tau BK}}^{(+)}\left(k\right)$, which is a Hatano-Nelson model\,\citep{Hatano1996PRL}. In terms of a point gap, it belongs to class AIII and has a nontrivial winding number (\ref{eq:SM_Winding_Spec}) for $0\leq\left|\mu^{\prime}\right|<\left|\Delta^{\prime}\right|$. 

\subsection{Additional sublattice symmetry}
Now let us assume that there is an additional sublattice symmetry for the dynamical matrix, i.e., 
\begin{equation}
	{\cal S}H_{\tau}\left(k\right){\cal S}^{-1}=-H_{\tau}\left(k\right),\ \ {\cal S}^2=I,
\end{equation}
where ${\cal S}$ is unitary. In the presence of the sublattice symmetry ${\cal S}$, the QBS can also respects the chiral symmetry, which is the combination of the pseudo-Hermiticity and  sublattice symmetry, 
\begin{equation}
	\Sigma H_{\tau}\left(k\right)\Sigma^{-1}=-H_{\tau}^{\dagger}\left(k\right),\ \ \ \Sigma^2=I,
\end{equation}
where $\Sigma\coloneqq\eta{\cal S}\equiv\tau^{3}{\cal S}$ is unitary. Therefore, it gives rise to an unconventional time-reversal symmetry, 
\begin{equation}
	{\cal T}H_{\tau}^{T}\left(-k\right){\cal T}^{-1}=H_{\tau}\left(k\right), \ \ \ {\cal T}{\cal T}^{*}=-I,
	\label{eq:SM_TRS2}
\end{equation}
where ${\cal T}\coloneqq i\tau^{2}{\cal S}$. Note that ${\cal S}$ commutes with $\eta$ and ${\cal T}$, i.e., $\eta{\cal S}={\cal S}\eta$ and ${\cal T}{\cal S}={\cal S}{\cal T}$, for physical reasons. Due to the presence of the sublattice symmetry, the symmetry class for the QBSs becomes to class $\text{DIII}^\dagger$ given in Ref.\,\citep{Kawabata2019PRX}. The associated topological classification for 1D QBSs is $\mathbb{Z}_2$. The time-reversal symmetry~(\ref{eq:SM_TRS2}) supplies a $\mathbb{Z}_{2}$ invariant for the QBS. A concrete model is discussed in the main text. 

\section{Topological invariants in the squeezed SSH model \label{sec:SM_SSH}}
To derive the topological invariants of the squeezed SSH model in the main text, we start from the Hamiltonian of system in the crystal-momentum space $\hat{H}_{{\rm SSH}}=\frac{1}{2}\sum_{k}\hat{\Phi}_{k}^{\dagger}H_{{\rm SSH}}(k)\hat{\Phi}_{k}$, where $\hat{\Phi}_{k}=(\hat{a}_{kA},\hat{a}_{kB},\hat{a}_{-kA}^{\dagger},\hat{a}_{-kB}^{\dagger})^{T}$. Here the first-quantized Hamiltonian in the $k$-space can be expressed as 
\begin{equation}
	\begin{split}H_{{\rm SSH}}\left(k\right) & =\left(t_{1}+t_{2}\cos k\right)\tau^{0}\sigma^{1}+t_{2}\sin k\tau^{0}\sigma^{2}+\left(g_{1}+g_{2}\cos k\right)\tau^{1}\sigma^{1}+g_{2}\sin k\tau^{1}\sigma^{2},
	\end{split}
	\label{eq:SM_H_SSH}
\end{equation}
where $\sigma^{j}$ with $j=1,2,3$ are the Pauli matrices and $\tau^{1}$ is the particle hole or charge conjugation. Note that $\sigma^{0}$ and $\tau^{0}$ are identity matrices. From the eigenfunction $\det\left(H_{\tau{\rm SSH}}\left(k\right)-EI\right)=0$ with $H_{\tau{\rm SSH}}\left(k\right)\coloneqq\tau^{3}H_{{\rm SSH}}\left(k\right)$, we obtain the spectrum 
\begin{equation}
	E_{\pm}^{2}\left(k\right)=\Delta^{2}+2\left(t_{1}t_{2}-g_{1}g_{2}\right)\cos k\pm2 i\left(t_{1}g_{2}-t_{2}g_{1}\right)\sin k,
	\label{eq:SM_SSH_Spec}
\end{equation}
where $\Delta\coloneqq\sqrt{t_1^2+t_2^2-g_1^2-g_2^2}$ is real ($\mathbb{R}$) or purely imaginary ($i\mathbb{R}$).
The squeezed SSH model respects the sublattice symmetry 
\begin{equation}
	{\cal S}H_{\tau\rm SSH}\left(k\right){\cal S}^{-1}=-H_{\tau\rm SSH}\left(k\right),\label{eq:SM_SLS}
\end{equation}
with ${\cal S}\coloneqq \tau^0\sigma^{3}$, which enriches the topological phase of the model. 
\begin{figure}
	\begin{centering}
		\includegraphics[width=17cm]{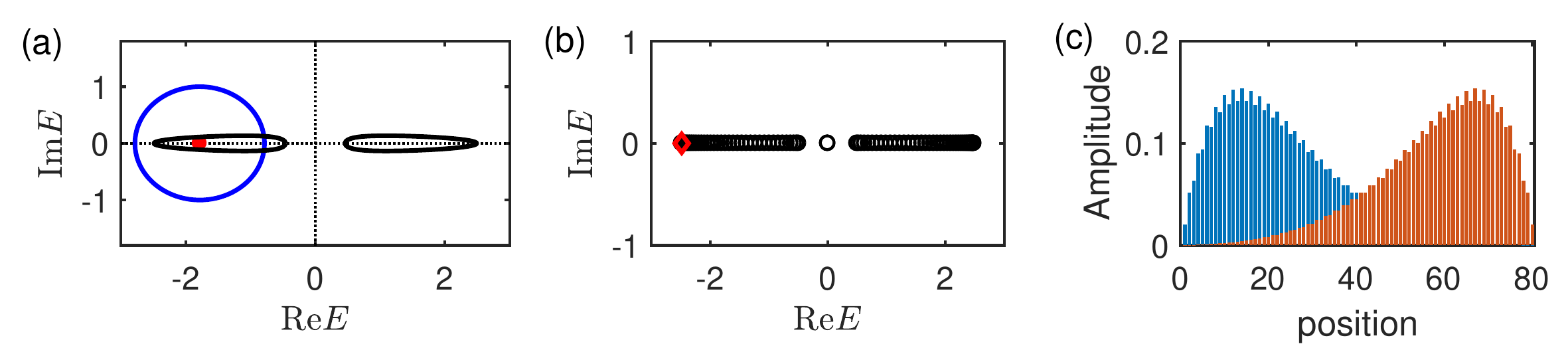} 
		\par\end{centering}
	\caption{(a,b) Spectra (black) of the squeezed SSH model in the PBC (a) and OBC (b). The blue unit circle in (a) denotes the spectrum of the unitary matrix $U_{\tau \rm SSH}(k)$ for $E=-\Delta$ (red dot). (c) Amplitude of the Kramers pair with the lowest energy {[}marked by a red diamond in (b){]}, indicating the Kramers degeneracy. Other parameters are $t_{2}=3t_{1}/2$, $g_{1}=0$, $g_{2}=0.2t_{1}$ and $L=40$. 
		\label{fig_S1}}
\end{figure}
\subsection{Point-gap topology}
In terms of a point gap at the reference energy $E$, i.e., $\det(H_{\tau\rm SSH}(k)-EI)\neq0$, $H_{\tau{\rm SSH}}\left(k\right)$ can be continuously deformed to a unitary matrix without closing the gap 
\begin{equation}
	H_{{\rm \tau SSH},\lambda}\left(k\right)\coloneqq\left(1-\lambda\right)H_{\tau{\rm SSH}}\left(k\right)+\lambda\left(U_{\tau{\rm SSH}}\left(k\right)+EI\right),
\end{equation}
with $U_{\tau{\rm SSH}}\left(k\right)\coloneqq[\left(H_{\tau{\rm SSH}}\left(k\right)-EI\right)\left(H_{\tau{\rm SSH}}\left(k\right)-EI\right)^{\dagger}]^{-1/2}\left(H_{\tau{\rm SSH}}\left(k\right)-EI\right)$ and $\lambda\in[0,1]$.
The squeezed SSH model respects an unconventional time-reversal symmetry in the presence of the sublattice symmetry (\ref{eq:SM_SLS}),
\begin{equation}
	\begin{split}{\cal T}H_{\tau{\rm SSH}}^{T}\left(-k\right){\cal T}^{-1} & =H_{\tau{\rm SSH}}\left(k\right),\end{split}
	\label{eq:SM_TRS}
\end{equation}
where ${\cal T}\coloneqq i\tau^{2}\sigma^{3}$ (${\cal T}{\cal T}^{*}=-I$). The time-reversal symmetry (\ref{eq:SM_TRS}) consequently yields a twofold degeneracy of the complex spectrum, which is a non-Hermitian generalization of Kramers' theorem. 

This model belongs to class $\text{D}^{\dagger}$ with respect to the reference energy $E$ and the time-reversal symmetry (\ref{eq:SM_TRS}) supplies a $\mathbb{Z}_{2}$ invariant although the winding number (\ref{eq:SM_Winding_Spec}) is always vanishing. In terms of the Pfaffian, this $\mathbb{Z}_{2}$ invariant $\nu\left(E\right)$ is defined as 
\begin{equation}
	\left(-1\right)^{\nu\left(E\right)}\coloneqq{\rm sgn}\left\{ \frac{{\rm Pf}\left(H_{\tau{\rm SSH}}\left(0\right){\cal T}-E{\cal T}\right)}{{\rm Pf}\left(H_{\tau{\rm SSH}}\left(\pi\right){\cal T}-E{\cal T}\right)}\exp\left[-\frac{1}{2}\int_{0}^{\pi}dk\frac{\partial}{\partial k}\ln\det\left(H_{\tau{\rm SSH}}\left(k\right){\cal T}-E{\cal T}\right)\right]\right\} .\label{eq:SM_Z2_ind}
\end{equation}
Due to the particle-hole symmetry, the exponential in Eq.\,(\ref{eq:SM_Z2_ind}) equals to the identity for $E=0$. Thus, the definition of this invariant $\nu\left(E\right)$ at zero energy is consistent with Eq.\,(5) in
the main text. The $\mathbb{Z}_{2}$ index $\nu\left(E=0\right)\in\left\{ 0,1\right\}$ is nontrivial only if the relation 
\begin{equation}
	\left(\Delta^{2}-E^{2}\right)^{2}<4\left(t_{1}t_{2}-g_{1}g_{2}\right)^{2}
\end{equation}
is satisfied. Note that $\pm\Delta$ is in the area enclosed by the spectrum of the model. Corresponding to the main text, Figure\,\ref{fig_S1}(a) shows that it is always possible to continuously deform $E$ to $\Delta$ or $-\Delta$ without closing the point gap if $E$ is in the loop of the spectrum. Hence, the $\mathbb{Z}_{2}$ invariant $\nu\left(E\right)$ in such a case is nontrivial under the condition of $t_{1}t_{2}\neq g_{1}g_{2}$.

\subsection{Line-gap topology}
Now we discuss the line-gap topology of the squeezed model. As mentioned in the above, $\pm\Delta$ is inside the area encircled by the spectrum of system. The real (imaginary) gap ${\rm Re}E=0$ (${\rm Im}E=0$) opens only if $\pm\Delta$ cannot be deformed to the reference energy $E=0$ without closing the point gap. In this case, the point-gap topology of the system is trivial at zero energy, i.e., $\nu\left(E=0\right)=0$. Therefore, from Eq.\,(\ref{eq:SM_Z2_ind}), we obtain the gap-opening conditions for the real and imaginary gaps, i.e., $\Delta^{2}>2\left|t_{1}t_{2}-g_{1}g_{2}\right|$ and $\Delta^{2}<-2\left|t_{1}t_{2}-g_{1}g_{2}\right|$, respectively. 

Before defining the topological invariant, we notice two homotopic equivalences in terms of a line gap. The dynamical matrix $H_{\tau{\rm SSH}}\left(k\right)$ can be continuously deformed to a Hermitian matrix for the real gap and an anti-Hermitian one for the imaginary gap. It implies that the real-gap (imaginary-gap) topology is determined by the particle-conserving part (two-mode squeezing term). Therefore, for simplification, we can focus on the following Hermitian and anti-Hermitian matrices for the real and imaginary gaps, respectively, i.e.,
\begin{equation}
	\begin{split}H_{\tau{\rm SSH}}^{\left({\rm real}\right)}\left(k\right) & =\frac{H_{\tau{\rm SSH}}\left(k\right)+H_{\tau{\rm SSH}}^{\dagger}\left(k\right)}{2}=\left[\left(t_{1}+t_{2}\cos k\right)\sigma^{1}+t_{2}\sin k\sigma^{2}\right]\tau^{3},
	\end{split}
	\label{eq:Re_Top}
\end{equation}
and 
\begin{equation}
	\begin{split}H_{\tau{\rm SSH}}^{{\rm \left(imaginary\right)}}\left(k\right) & =\frac{H_{\tau{\rm SSH}}\left(k\right)-H_{\tau{\rm SSH}}^{\dagger}\left(k\right)}{2 i}=\left[\left(g_{1}+g_{2}\cos k\right)\sigma^{1}+g_{2}\sin k\sigma^{2}\right]\tau^{2}.
	\end{split}
	\label{eq:Im_Top}
\end{equation}
From Eqs.\,(\ref{eq:Re_Top}) and (\ref{eq:Im_Top}), the topological phase of
the squeezed SSH model are characterized by the winding numbers $W^{\left({\rm real}\right)}$ and $W^{\left({\rm imaginary}\right)}$ for the real and imarginary gaps, respectively. When the real ($\Delta^{2}>2\left|t_{1}t_{2}-g_{1}g_{2}\right|$) and imaginary ($\Delta^{2}<-2\left|t_{1}t_{2}-g_{1}g_{2}\right|$) gaps open, the winding numbers are given by 
\begin{equation}
	W^{\left({\rm real}\right)}=\int_{{\rm BZ}}\left(q\left(k\right)\right)^{-1}dq\left(k\right)=
	\begin{cases}
		1, & t_{1}<t_{2}\\
		0, & t_{1}>t_{2}
	\end{cases}\label{W_real}
\end{equation}
with $q\left(k\right)=t_{1}+t_{2}e^{ ik}$, and 
\begin{equation}
	W^{\left({\rm imaginary}\right)}=\int_{{\rm BZ}}\left(p\left(k\right)\right)^{-1}dp\left(k\right)=
	\begin{cases}
		1, & \left|g_{1}\right|<\left|g_{2}\right|\\
		0, & \left|g_{1}\right|>\left|g_{2}\right|
	\end{cases}\label{W_imaginary}
\end{equation}
with $p\left(k\right)=g_{1}+g_{2}e^{ ik}$, respectively. 
%As shown in Fig.\,3 of the main text, the winding number for the real gap is nonvanishing, i.e., $W^{\left({\rm real}\right)}=1$, in the region $t_{1}+\left|g_{2}\right|<t_{2}$ for $g_{1}=0$.

\section{Skin effect and zero modes in the squeezed SSH model }\label{sec:SM_SkinEffect}
In this section, we will show the detailed derivation for the reported skin effect and zero modes to support the main text. Specifically, by imposing the OBC to the system, the Hamiltonian reads $\hat{H}_{{\rm SSH}}=\frac{1}{2}\hat{\Phi}^{\dagger}H_{{\rm SSH}}\hat{\Phi}$, where $\hat{\Phi}=(\hat{a}_{1A},\hat{a}_{1B},\ldots,\hat{a}_{LA},\hat{a}_{LB},\hat{a}_{1A}^{\dagger},\ldots,\hat{a}_{LB}^{\dagger})^{T}$.
Here the first-quantized Hamiltonian can be expressed by the $4L$-by-$4L$ matrix 
\begin{equation}
	\begin{split}H_{{\rm SSH}} & =t_{1}\tau^0\sigma^{1}I+\frac{t_{2}}{2}\tau^0\sigma^{1}\left(T+T^{\dagger}\right)+\frac{t_{2}}{2i}\tau^0\sigma^{2}\left(T-T^{\dagger}\right)+g_{1}\tau^{1}\sigma^{1}I+\frac{g_{2}}{2}\tau^{1}\sigma^{1}\left(T+T^{\dagger}\right)+\frac{g_{2}}{2i}\tau^{1}\sigma^{2}\left(T-T^{\dagger}\right),
	\end{split}
	\label{eq:SM_H_OBC}
\end{equation}
where $T=\left(\delta_{j, j'-1}\right)$ is the $L$-dimensional Toeplitz matrix  ($j,j'=1,2,..L$). Here $L$ is the cell number of SSH model.

\subsection{Relation between the squeezing transformation and skin effect}
It is predicted by Eq.\,(\ref{eq:SM_Z2_ind}) that there exists a skin effect in the QBS under the OBC (corresponding to the nontrivial $\mathbb{Z}_{2}$ index), if the reference energy $E$ is in the loop of the spectrum, e.g., $E=\pm\Delta$. To establish the relation between the squeezing transformation and skin effect, we introduce the squeezing operator for the squeezed SSH model under the OBC \citep{Wan2021PRA} 
\begin{equation}
	\hat{S}=\exp\left(\frac{1}{2}\hat{\Phi}^{\dagger}\tau^{3}W\hat{\Phi}\right),
\end{equation}
where the undetermined matrix $W$ is Hermitian and obeys $\eta W\eta^{-1}=-W$ with $\eta=\tau^{3}\sigma^0 I$ and ${\cal C}W^{*}{\cal C}^{-1}=W$ with ${\cal C}=\tau^{1}\sigma^0 I$. Note
that $\hat{S}$ is unitary, i.e., $\hat{S}^{-1}=\hat{S}^{\dagger}$. By performing the squeezing transformation on the Nambu spinor $\hat{\Phi}$, we obtain 
\begin{equation}
	\hat{S}^{\dagger}\hat{\Phi}\hat{S}=e^{W}\hat{\Phi}.
	\label{eq:SM_Sqz1}
\end{equation}
Substituting Eq.\,(\ref{eq:SM_Sqz1}) to the Hamiltonian $\hat{H}_{{\rm SSH}}=\hat{\Phi}^{\dagger}H_{{\rm SSH}}\hat{\Phi}/2$ of the squeezed SSH model under the OBC, we obtain  
\begin{equation}
	\hat{S}^{\dagger}\hat{H}_{{\rm SSH}}\hat{S}=\frac{1}{2}\hat{\Phi}^{\dagger}e^{W}H_{{\rm SSH}}e^{W}\hat{\Phi}=\frac{1}{2}\hat{\Phi}^{\dagger}H_{{\rm SSH}}'\hat{\Phi}.
	\label{eq:SM_Sqz_SQ}
\end{equation}
From Eq.\,(\ref{eq:SM_Sqz_SQ}), thus, the squeezing transformation on the first-quantized Hamiltonian can be rewritten as 
\begin{equation}
	H_{{\rm SSH}}^{\prime}=e^{W}H_{{\rm SSH}}e^{W}.
	\label{eq:SM_Sqz2}
\end{equation}
Notably, $e^{W}$ satisfying $e^{-W}=\eta e^{W}\eta^{-1}$ and $\det e^{W}=1$ is one entry of the special pseudounitary group $e^{W}\in{\rm SU}\left(2L,2L\right)$. Therefore, the squeezing transformation is unitary in the second-quantization language, and it belongs to ${\rm SU}\left(2L,2L\right)$ in the first-quantization language, as stated in the main text. 

For the dynamical stability, we assume $\left|g_{1}\right|<t_{1}$ and $\left|g_{2}\right|<t_{2}$. Setting $W=W_{1}+W_{2}$ where $W_{1}\coloneqq-r_{1}\tau^{1}\sigma^{0}I$ and $W_{2}\coloneqq-r_{2}\tau^{1}\sigma^{3}{\cal L}$ with the parameters $\tanh2r_{1}=g_{1}/t_{1}$, $\tanh r_{2}=(t_{1}g_{2}-t_{2}g_{1})/(t_{1}t_{2}-g_{1}g_{2})$ and ${\cal L}={\rm diag}(1,2,\ldots,L)$, we obtain 
\begin{equation}
	\begin{split}H_{{\rm SSH}}^{\prime}=e^{W}H_{{\rm SSH}}e^{W} & =\left(t_{1}\tau^{0}+g_{1}\tau^{1}\right)e^{-2r_{1}\tau^{1}}\sigma^{1}I+\left(t_{2}\tau^{0}+g_{2}\tau^{1}\right)e^{-2r_{1}\tau^{1}-r_{2}\tau^{1}}\left(\sigma^{-}T^{\dagger}+{\rm H.c.}\right)\\
		& =\tau^{0}\left[\tilde{t}_{1}\sigma^{1}I+\tilde{t}_{2}\left(\sigma^{-}T^{\dagger}+\sigma^{+}T\right)\right]\\
		&=h_{\rm SSH}\oplus h_{\rm SSH},
	\end{split}
	\label{eq:SM_Sqz_SSH}
\end{equation}
where $\tilde{t}_{1}\coloneqq\sqrt{t_{1}^{2}-g_{1}^{2}}$ and $\tilde{t}_{2}\coloneqq\sqrt{t_{2}^{2}-g_{2}^{2}}$. The properties $\left[{\cal L},T\right]=T$, $\left[{\cal L},T^{\dagger}\right]=-T$ and $\sigma^{3}\sigma^{\pm}=\pm\sigma^{\pm}$ with $\sigma^{\pm}\coloneqq\left(\sigma^{1}\pm i\sigma^{2}\right)/2$ have been used here. Note that Eq.\,(\ref{eq:SM_Sqz_SSH}) describes the conventional SSH model, which is consistent with the mapped Hamiltonian of the main text. It implies that the skin effect manifested by $\exp W_{2}$ corresponds to the two-mode squeezing in the squeezed SSH model. In the continuum limit $L\rightarrow\infty$, the continuum bands of Eq.\,(\ref{eq:SM_Sqz_SSH}) are given by 
\begin{equation}
	E=\pm\sqrt{\tilde{t}_{1}^{2}+\tilde{t}_{2}^{2}+2\tilde{t}_{1}\tilde{t}_{2}\cos k},\label{eq:SM_SSH_Spec_Deform}
\end{equation}
with the quasi-momentum $k\in\left[-\pi,\pi\right)$. These continuum bands are plotted by the red lines in Fig.\,2(a) of the main text. 

In the dynamically unstable regime, the dynamical matrix of the squeezed SSH model still can be transformed to a normal matrix (i.e., the skin effect disappearing) by the squeezing transformation. In terms of the squeezing parameters $r_{1}$ and
$r_{2}$, we summary it into three types of instability as follows. 
\begin{itemize}
	\item Type I: We assume $\tanh2r_{1}=t_{1}/g_{1}$ and $\tanh\left(2r_{1}+r_{2}\right)=t_{2}/g_{2}$ for $t_{1,2}<\left|g_{1,2}\right|$. Performing the squeezing transformation (\ref{eq:SM_Sqz1}), the squeezed Hamiltonian is given by 
	\begin{equation}
		\hat{H}_{{\rm SSH}}^{\left({\rm I}\right)}=\sum_{j=1}^{L}\left(\tilde{g}_{1}\hat{a}_{jA}^{\dagger}\hat{a}_{jB}^{\dagger}+\tilde{g}_{2}\hat{a}_{j+1A}^{\dagger}\hat{a}_{jB}^{\dagger}+{\rm H.c.}\right),
		\label{eq:SM_Instability1}
	\end{equation}
	where $\tilde{g}_{1}\coloneqq\sqrt{g_{1}^{2}-t_{1}^{2}}$ and $\tilde{g}_{2}\coloneqq\sqrt{g_{2}^{2}-t_{2}^{2}}$.  
	\item Type II: Similar with Type I, we take $\tanh2r_{1}=g_{1}/t_{1}$ and $\tanh\left(2r_{1}+r_{2}\right)=t_{2}/g_{2}$ for $t_{1}>\left|g_{1}\right|$ and $t_{2}<\left|g_{2}\right|$. After some algebra, the squeezed
	Hamiltonian is obtained, 
	\begin{equation}
		\hat{H}_{{\rm SSH}}^{\left({\rm II}\right)}=\sum_{j=1}^{L}\left(\tilde{t}_{1}\hat{a}_{jA}^{\dagger}\hat{a}_{jB}+\tilde{g}_{2}\hat{a}_{j+1A}^{\dagger}\hat{a}_{jB}^{\dagger}+{\rm H.c.}\right).
		\label{eq:SM_Instability2}
	\end{equation}
	\item Type III: Taking $\tanh2r_{1}=t_{1}/g_{1}$ and $\tanh\left(2r_{1}+r_{2}\right)=g_{2}/t_{2}$ for $t_{1}<\left|g_{1}\right|$ and $t_{2}>\left|g_{2}\right|$, the corresponding squeezed Hamiltonian is given by 
	\begin{equation}
		\hat{H}_{{\rm SSH}}^{\left({\rm III}\right)}=\sum_{j=1}^{L}\left(\tilde{g}_{1}\hat{a}_{jA}^{\dagger}\hat{a}_{jB}^{\dagger}+\tilde{t}_{2}\hat{a}_{j+1A}^{\dagger}\hat{a}_{jB}+{\rm H.c.}\right).
		\label{eq:SM_Instability3}
	\end{equation}
\end{itemize}
It is shown from Eqs.\,(\ref{eq:SM_Instability1}-\ref{eq:SM_Instability3}) that the correspondence between the skin effect and squeezing transformation still exist in the dynamically unstable regime. 

\subsection{Skin effect induced by squeezing}
Next we intend to derive the Kramers pair manifesting the $\mathbb{Z}_{2}$ skin effect. Firstly, from Eq.\,(\ref{eq:SM_Sqz_SSH}), we obtain the eigenequation 
\begin{equation}
	H_{\tau{\rm SSH}}e^{W}u^{\left(+\right)}=Ee^{W}u^{\left(+\right)},\ \ \ H_{{\rm \tau SSH}}e^{W}{\cal S}u^{\left(-\right)}=Ee^{W}{\cal S}u^{\left(-\right)},\label{eq:SM_EigenEq}
\end{equation}
where $E\in\mathbb{R}$ is the energy of system, and $u^{(+)}=\left(u^{T},0\right)^{T}$ and $u^{(-)}=(0,u^{\dagger})^{T}$ with $u=\left(u_{1A},u_{1B},\ldots,u_{LA},u_{LB}\right)^{T}$. $u$ is the  the eigenstate of $h_{\rm SSH}$ with energy $E$. For convenience, $g_{1}=0$ and $t_{2}>\left|g_{2}\right|$ are assumed. We can obtain the Kramers degeneracy by taking the superposition of the particle and hole $(v_{+},v_{-})$ with $v_{+}\coloneqq\left(e^{W}u^{(+)}+e^{W}{\cal S}u^{\left(-\right)}\right)/\sqrt{2}$ and $v_{-}\coloneqq(e^{W}u^{(+)}-e^{W}{\cal S}u^{(-)})/\sqrt{2}i$. In the second-quantization language, these states are rewritten as
\begin{equation}
	\begin{split}\hat{\phi}_{+} & \coloneqq v_{+}^{T}\hat{\Phi}=\sum_{j=1}^{L}e^{-r_{2}j}\left(u_{jA}\hat{x}_{jA}+ iu_{jB}\hat{p}_{jB}\right),\\
		\hat{\phi}_{-} & \coloneqq v_{-}^{T}\hat{\Phi}=\sum_{j=1}^{L}e^{r_{2}j}\left(u_{jA}\hat{p}_{jA}- iu_{jB}\hat{x}_{jB}\right),
	\end{split}
	\label{eq:SM_Skin_Mode1}
\end{equation}
where $\hat{x}_{j,\sigma}=(\hat{a}_{j,\sigma}+\hat{a}_{j,\sigma}^{\dagger})/\sqrt{2}$ and $\hat{p}_{j,\sigma}=(\hat{a}_{j,\sigma}-\hat{a}_{j,\sigma}^{\dagger})/\sqrt{2}i$ ($\sigma=A,B$) are the canonical coordinates and momenta, respectively. The above two modes form a Kramers pair since $[\hat{\phi}_{+},\hat{\phi}_{-}]=0$, and $\hat{\phi}^{(+)}$ ($\hat{\phi}^{(-)}$) is localized at left (right) if $g_{2}>0$ (i.e., $r_{2}>0$). Otherwise, the localization changes. After organizing, we then obtain the localized Kramers pair with energy $E$, which is given by 
\begin{equation}
	\begin{split}\hat{\phi}_{\text{L}} & =\sum_{j=1}^{L}\left(e^{-\left|r_{2}\right|}\right)^{j-1}\left(u_{jA}\hat{x}_{jA}+ isu_{jB}\hat{p}_{jB}\right),\\
		\hat{\phi}_{\text{R}} & =\sum_{j=1}^{L}\left(e^{-\left|r_{2}\right|}\right)^{L-j}\left(u_{jA}\hat{p}_{jA}- isu_{jB}\hat{x}_{jB}\right),
	\end{split}
	\label{eq:SM_Skin_Mode2}
\end{equation}
with $s={\rm sgn}(g_2)=\pm$. Here the subscripts ``L'' and ``R'' denote the left and right, respectively. This localization of the Kramers pair manifesting the $\mathbb{Z}_{2}$ skin effect is numerically plotted in Fig.\,2(b) of the main text. 

\subsection{Zero modes}\label{subsec:ZM}
Now we derive the zero modes in the dynamical-stability regime. Consistent with the main text, the mapped Hamiltonian describing the conventional SSH model under the OBC reads 
\begin{equation}
	\begin{split}\hat{H}_{{\rm SSH}} & =\sum_{j=1}^{L}\tilde{t}_{1}\hat{\alpha}_{jA}^{\dagger}\hat{\alpha}_{jB}+\tilde{t}_{2}\hat{\alpha}_{j+1A}^{\dagger}\hat{\alpha}_{jB}+{\rm H.c.}\\
		& =\hat{\bm{\alpha}}^{\dagger}h_{{\rm SSH}}\hat{\bm{\alpha}},
	\end{split}
\end{equation}
where $\hat{\bm{\alpha}}=\left(\hat{\alpha}_{1A},\hat{\alpha}_{1B},\ldots,\hat{\alpha}_{LA},\hat{\alpha}_{LB}\right)^{T}$ satisfying 
\begin{equation}
	\left(\begin{array}{c}
		\hat{\alpha}_{jA}\\
		\hat{\alpha}_{jA}^{\dagger}
	\end{array}\right)=e^{\left(r_{1}+r_{2}j\right)\tau^{1}}\left(\begin{array}{c}
		\hat{a}_{jA}\\
		\hat{a}_{jA}^{\dagger}
	\end{array}\right),\ \ \ \left(\begin{array}{c}
		\hat{\alpha}_{jB}\\
		\hat{\alpha}_{jB}^{\dagger}
	\end{array}\right)=e^{\left(r_{1}-r_{2}j\right)\tau^{1}}\left(\begin{array}{c}
		\hat{a}_{jB}\\
		\hat{a}_{jB}^{\dagger}
	\end{array}\right).
\end{equation}
Here $h_{{\rm SSH}}=\tilde{t}_{1}\sigma^{1}I_{L}+\tilde{t}_{2}\left(\sigma^{-}T^{\dagger}+\sigma^{+}T\right)$ is the first-quantized Hamiltonian of the SSH model. From the eigenequation $h_{{\rm SSH}}{\rm ZM}_{\text{L}/\text{R}}=0$, we obtain the zero modes 
\begin{equation}
	\begin{split}{\rm ZM}_{\text{L}} & =\left(1,0,\delta,0,\ldots,\delta^{L-1},0\right)^{T},\\
		{\rm ZM}_{\text{R}} & =\left(0,\delta^{L-1},0,\delta^{L-2},\ldots,0,1\right)^{T},
	\end{split}
\end{equation}
with $\delta\coloneqq-\tilde{t}_{1}/\tilde{t}_{2}$ ($\left|\delta\right|<1$). Here $\delta$ is the localization parameter of zero modes in the conventional SSH model. In the canonical coordinates and momenta representation, $\hat{X}_{j\sigma}\coloneqq (\hat{\alpha}_{j\sigma}+\hat{\alpha}_{j\sigma}^\dagger)/\sqrt{2}$ and $\hat{P}_{j\sigma}\coloneqq (\hat{\alpha}_{j\sigma}-\hat{\alpha}_{j\sigma}^\dagger)/\sqrt{2i}$, we then obtain four zero modes 
\begin{align}
	&\hat{\rm X}_{\rm L}=\sum_{j=1}^L \frac{{\rm ZM}_{\rm L}^T\hat{\alpha}+\hat{\alpha}^\dagger{\rm ZM}_{\rm L}}{\sqrt{2}}, \ \  \hat{\rm X}_{\rm R}=\sum_{j=1}^L \frac{{\rm ZM}_{\rm R}^T\hat{\alpha}+\hat{\alpha}^\dagger{\rm ZM}_{\rm R}}{\sqrt{2}}, \notag \\
	&\hat{\rm P}_{\rm L}=\sum_{j=1}^L \frac{{\rm ZM}_{\rm L}^T\hat{\alpha}-\hat{\alpha}^\dagger{\rm ZM}_{\rm L}}{\sqrt{2i}}, \ \  \hat{\rm P}_{\rm R}=\sum_{j=1}^L \frac{{\rm ZM}_{\rm R}^T\hat{\alpha}-\hat{\alpha}^\dagger{\rm ZM}_{\rm R}}{\sqrt{2i}}.
\end{align}
Using the derived Kramers pair~(\ref{eq:SM_Skin_Mode1}) for $g_{1}=0$, we immediately obtain the zero modes in the original basis, and they are expressed as 
\begin{equation}
	\begin{split}\hat{x}_{{\rm L}}=\sum_{j=1}^{L}\delta^{j-1}e^{r_{2}j}\hat{x}_{jA}, & \ \ \ \hat{x}_{{\rm R}}=\sum_{j=1}^{L}\delta^{L-j}e^{-r_{2}j}\hat{x}_{jB},\\
		\hat{p}_{{\rm L}}=\sum_{j=1}^{L}\delta^{j-1}e^{-r_{2}j}\hat{p}_{jA}, & \ \ \ \hat{p}_{{\rm R}}=\sum_{j=1}^{L}\delta^{L-j}e^{r_{2}j}\hat{p}_{jB}.
	\end{split}
	\label{eq:SM_Sqz_ZM}
\end{equation}
From Eq.\,(\ref{eq:SM_Sqz_ZM}), it yields the relation 
\begin{equation}
	\delta_{\pm s}\coloneqq\frac{-t_{1}}{t_{2}\pm s\left|g_{2}\right|}=\delta e^{\mp s\left|r_{2}\right|}.
	\label{eq:SM_Localization}
\end{equation}
Such a derived localization parameter (\ref{eq:SM_Localization}) for the squeezed SSH model determines the occurrence of zero modes in two topological phases: (i) the (conventional) real-gap topological and (ii) real- and point-gap coexisting topological phase. The former characterized by the nonvanishing winding number $W^{\left({\rm real}\right)}$ corresponds to the case of the real gap opening under the PBC. And four zero modes occur under the OBC since $|\delta_{\pm}|<1$ is satisfied. In the real- and point-gap coexisting topological phase, the real-gap topology survives while the real gap is closed. Meanwhile, the point-gap topology is nontrivial in this coexisting phase. It indicates an anomalous bulk-boundary correspondence, and there are two zero modes appearing under the OBC due to $|\delta_{+}|<1$ and $|\delta_{-}|>1$.

Specifically, the real-gap topological phase, i.e., $t_{2}-\left|g_{2}\right|>t_{1}$ (yellow area in Fig.\,3 of the main text), corresponds to $\left|\delta_{+}\right|<\left|\delta_{-}\right|<1$. Then we obtain two pairs of the zero modes as shown in the main text
\begin{equation}
	\begin{split}\hat{x}_{{\rm L}}^{-s}=\sum_{j=1}^{L}\delta_{-s}^{j-1}\hat{x}_{jA}, & \ \ \ \hat{x}_{{\rm R}}^{-s}=\sum_{j=1}^{L}\delta_{-s}^{L-j}\hat{x}_{jB},\\
		\hat{p}_{{\rm L}}^{s}=\sum_{j=1}^{L}\delta_{s}^{j-1}\hat{p}_{jA}, & \ \ \ 
		\hat{p}_{{\rm R}}^{s}=\sum_{j=1}^{L}\delta_{s}^{L-j}\hat{p}_{jB}.
	\end{split}\label{eq:SM_Sqz_ZM2}
\end{equation}
From the commutation relations $[\hat{x}_{{\rm L/R}}^{-s},\hat{H}_{{\rm SSH}}]=it_{1}\delta_{-s}^{N-1}\hat{p}_{LB/1A}$ and $[\hat{p}_{\text{L/R}}^{s},\hat{H}_{{\rm SSH}}]=-it_{1}\delta_{s}^{N-1}\hat{x}_{LB/1A}$, these modes are approximately conserved. Note that the chirality makes the left and right boundary states are commutative, i.e., $[\hat{x}_{\text{L}}^{-s},\hat{p}_{\text{R}}^{s}]=[\hat{x}_{\text{R}}^{-s},\hat{p}_{\text{L}}^{s}]=0$. And the paired zero modes satisfy $[\hat{x}_{\text{L}}^{-s},\hat{p}_{\text{L}}^{s}]=[\hat{x}_{\text{R}}^{-s},\hat{p}_{\text{R}}^{s}]= i\left(1-\delta^{2L}\right)/\left(1-\delta^{2}\right)$, which indicates that they are canonical conjugated quantities. 

Equation (\ref{eq:SM_Localization}) also shows the existence of $\hat{x}^{-}_{\rm L/R}$ or $\hat{p}^{-}_{\rm L/R}$ is determined by the localization competition between the skin effect ($e^{|r_2|}>1$) and the zero modes of the conventional SSH model ${\rm ZM}_{\rm L/R}$ ($|\delta|<1$). Once the skin effect dominates, i.e., $|\delta_{-}|=e^{|r_2|}|\delta|>1$, the occurrence of $\hat{x}^{-}_{\rm L/R}$ or $\hat{p}^{-}_{\rm L/R}$ is inhibited. For the coexisting topological phase (green area in Fig.~3 of the main text), the parameters satisfy the relation $0<t_{2}-\left|g_{2}\right|<t_{1}$ and $t_{2}>t_{1}$, corresponding to $\left|\delta_{+}\right|<1<\left|\delta_{-}\right|$. In this case, a pair of zero modes $\hat{x}^{-}_{\rm L/R}$ is inhibited by the skin effect for $g_2>0$. Meanwhile, the other zero modes $\hat{p}^{+}_{\rm L/R}$ survive due to $|\delta_{+}|<1$. Similarly, for $g_{2}<0$, one can only obtain a pair of zero modes consisting of canonical coordinates
\begin{equation}
	\begin{split}\hat{x}_{\text{L}}^{+}=\sum_{j=1}^{L}\delta_{+}^{j-1}\hat{x}_{jA}, & \ \ \ \hat{x}_{\text{R}}^{+}=\sum_{j=1}^{L}\delta_{+}^{L-j}\hat{x}_{jB},\end{split}
\end{equation}
and the zero modes $(\hat{p}^{-}_{\rm L}, \hat{p}^{-}_{\rm R})$ are inhibited. 
\begin{figure}
	\begin{centering}
		\includegraphics[width=16cm]{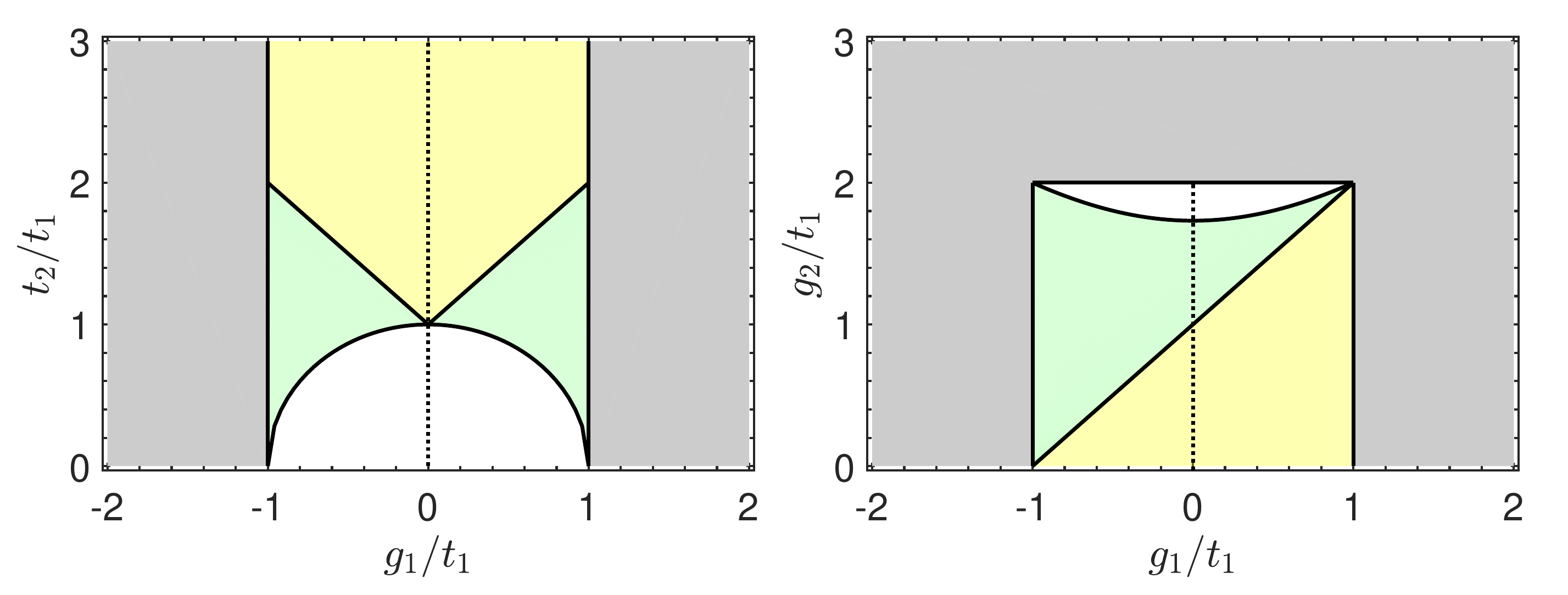} 
		\par\end{centering}
	\caption{Phase diagrams of the squeezed SSH model. The yellow area represents that the system has two pairs of zero modes under the OBC, indicating the conventional bulk-boundary correspondence. The green area represents the real- and point-gap coexisting topological phase, in which both the point-gap topology and real-gap topology are nontrivial and there is one pair of zero modes under the OBC. The system is topologically trivial in the white area and dynamically unstable in the gray area. Left panel: $g_{2}=0$. Right panel: $t_{2}=2t_{1}$. \label{fig_S2}}
\end{figure}

\subsection{Enriched phase diagram by the intracell squeezing}
The presence of the intra-cell squeezing with strength $g_{1}\neq0$ could enrich the phase diagram. We present a description for the phase diagram explicitly. 
\begin{itemize}
	\item The system is dynamically stable if $t_{1}>\left|g_{1}\right|$ and $t_{2}>\left|g_{2}\right|$. 
	\item The real gap of the original Hamiltonian in the main text opens if $t_{1}^{2}+t_{2}^{2}-g_{1}^{2}-g_{2}^{2}>2\left|t_{1}t_{2}-g_{1}g_{2}\right|$. The winding number is nontrivial if $t_{1}<t_{2}$, and the conventional
	bulk-boundary correspondence implies the occurrence of two pairs of zero modes. 
	\item The real gap of the mapped Hamiltonian in the main text opens if $\sqrt{t_{1}^{2}-g_{1}^{2}}\neq\sqrt{t_{2}^{2}-g_{2}^{2}}$. The winding number is nontrivial if $\sqrt{t_{1}^{2}-g_{1}^{2}}<\sqrt{t_{2}^{2}-g_{2}^{2}}$.
	In this regime, a pair of zero modes occurs according to the anomalous bulk-boundary correspondence. If the condition is not satisfied, there is no zero mode. 
\end{itemize}
Combining with them, one can obtain the whole phase diagram. In Fig.\,\ref{fig_S2}, we plot two phase diagrams enriched by the intra-cell squeezing, which is similar as Fig.\,3 of the main text, indicating an anomalous bulk-boundary correspondence.

\section{Reconstructing real-gap topology and infinitesimal instability of the skin effect\label{sec:SM_NonBloch}}
\subsection{Reconstruction of real-gap topology }
To fully characterize the real-gap topology in the dynamical-stability regime, here we perform the generalized Brillouin zone treatment\,\citep{Yokomizo2019PRL,Kawabata2020PRB,Yokomizo2021PRB}
on the squeezed SSH model, which is motivated by the anomalous bulk-boundary correspondence. From Eq.\,(\ref{eq:SM_SSH_Spec}) by taking $\beta\coloneqq e^{ ik}$, we obtain two decoupled equations 
\begin{align}
	E^{2}= & \left[t\left(\beta\right)-g\left(\beta\right)\right]\left[t\left(\beta^{-1}\right)+g\left(\beta^{-1}\right)\right],\label{eq:SM_Sol_Poly1}\\
	E^{2}= & \left[t\left(\beta\right)+g\left(\beta\right)\right]\left[t\left(\beta^{-1}\right)-g\left(\beta^{-1}\right)\right],\label{eq:SM_Sol_Poly2}
\end{align}
where $t\left(\beta\right)=t_{1}+t_{2}\beta$ and $g\left(\beta\right)=g_{1}+g_{2}\beta$. For a given $E$, the two algebraic equations are irreducible in terms of $\beta$. Remarkably, if $\beta$ satisfies Eq.\,(\ref{eq:SM_Sol_Poly1}), then $\beta^{-1}$ is a solution of Eq.\,(\ref{eq:SM_Sol_Poly2}), and vice versa. A fundamental solution with $\beta$ and another fundamental solution with $\beta^{-1}$ is linearly independent of each other, and they form a Kramers pair. Now, we define the solution of Eq.\,(\ref{eq:SM_Sol_Poly1}) as $\beta_{1}$ and $\beta_{2}$ with the order $\left|\beta_{1}\right|\leq\left|\beta_{2}\right|$, which satisfy
\begin{equation}
	\beta_{1}\beta_{2}=\frac{\left(t_{1}-g_{1}\right)\left(t_{2}+g_{2}\right)}{\left(t_{1}+g_{1}\right)\left(t_{2}-g_{2}\right)}.
\end{equation}
Then the solutions of Eq.\,(\ref{eq:SM_Sol_Poly2}) are given by $\beta_{1}^{-1}$ and $\beta_{2}^{-1}$. By performing the generalized Brillouin zone, we obtain 
\begin{equation}
	\left|\beta_{1}\right|=\left|\beta_{2}\right|=\sqrt{\left|\frac{\left(t_{1}-g_{1}\right)\left(t_{2}+g_{2}\right)}{\left(t_{1}+g_{1}\right)\left(t_{2}-g_{2}\right)}\right|}\equiv e^{r_{2}}\neq1,
	\label{eq:GBZ1}
\end{equation}
and the continuum bands are given by 
\begin{equation}
	E_{\pm}=\pm\sqrt{\Delta^{2}+2\sqrt{\left(t_{1}^{2}-g_{1}^{2}\right)\left(t_{2}^{2}-g_{2}^{2}\right)}\cos \tilde{k}}
	\label{eq:SM_Deform}
\end{equation}
for $\beta_{1}=\beta_{2}e^{ i \tilde{k}}$ with $\tilde{k}\in[-\pi,\pi)$, which is consistent with Eq.\,(\ref{eq:SM_SSH_Spec_Deform}). In particular, by the replacement $e^{ ik}I_{4}\rightarrow{\cal B}$ and $e^{- ik}I_{4}\rightarrow{\cal B}^{-1}$ with ${\cal B}=e^{ ikI_{4}-r_{2}\tau^{1}\sigma^{3}}$ for $g_{1}=0$ and $t_{2}>\left|g_{2}\right|$ , the dynamical matrix of system $H_{\tau \rm SSH}(k)$ becomes 
\begin{equation}
	\begin{split}H_{\tau{\rm SSH}}\left({\cal B}\right) & =t_{1}\tau^{3}\sigma^{1}+\left(t_{2}\tau^{3}+ ig_{2}\tau^{2}\right)\sigma^{+}{\cal B}^{-1}+\left(t_{2}\tau^{3}+ ig_{2}\tau^{2}\right)\sigma^{-}{\cal B}\\
		& =\left(\begin{array}{cc}
			t_{1}\sigma^{1}+\tilde{t}_{2}\left(\sigma^{+}e^{- ik}+\sigma^{-}e^{ ik}\right) & 0\\
			0 & -t_{1}\sigma^{1}-\tilde{t}_{2}\left(\sigma^{+}e^{- ik}+\sigma^{-}e^{ ik}\right)
		\end{array}\right)\\
		& \equiv\left(\begin{array}{cc}
			h_{{\rm SSH}}\left(k\right) & 0\\
			0 & -h_{{\rm SSH}}\left(k\right)
		\end{array}\right),
	\end{split}
	\label{eq:SM_Convention_SSH}
\end{equation}
where $\sigma^{\pm}=\left(\sigma^{1}\pm i\sigma^{2}\right)/2$ and $h_{{\rm SSH}}\left(k\right)=\left(t_{1}+\tilde{t}_{2}\cos k\right)\sigma^{1}+\tilde{t}_{2}\sin k\sigma^{2}$.
We have used the relation $\sigma^{\pm}\sigma^{3}=\mp\sigma^{\pm}$ in the above derivation. After the replacement, Equation~(\ref{eq:SM_Convention_SSH}) becomes the conventional SSH model in the PBC, which is consistent with Eq.\,(\ref{eq:SM_Sqz_SSH}). For the particle-conserving model $h_{{\rm SSH}}\left(k\right)$, the topology is characterized by the winding number 
\begin{equation}
	\tilde{W}^{\left({\rm real}\right)}=\frac{1}{2\pi i}\int_{{\rm BZ}}\left(\tilde{q}\left(k\right)\right)^{-1}d\tilde{q}\left(k\right)=\begin{cases}
		1, & t_{1}<\tilde{t}_{2}\\
		0, & t_{1}>\tilde{t}_{2}
	\end{cases},
\end{equation}
where $\tilde{q}\left(k\right)=t_{1}+\tilde{t}_{2}e^{ ik}$. 

\subsection{Infinitesimal instability of the skin effect against local perturbations}
The presence of the onsite perturbation $\hat{H}_{{\rm onsite}}=\mu\sum_{js}\hat{a}_{js}^{\dagger}\hat{a}_{js}$ breaks the time-reversal symmetry of the system, and thus the squeezed skin effect vanishes. To clearly show this, we focus on the eigenfunction of the squeezed model by setting $\beta\coloneqq e^{ ik}$, which is given by 
\begin{equation}
	\begin{split}\det\left(H_{\tau{\rm perturb}}\left(\beta\right)-EI\right) & =\det\left(H_{\tau{\rm SSH}}\left(\beta\right)-EI+\mu\tau^{3}\right)=0,
	\end{split}
\end{equation}
where $H_{\rm \tau perturb}=H_{\rm\tau SSH}(k)+\mu\tau^3$. We then arrive at a quartic equation  
\begin{equation}
	E^{4}-2[\mu^{2}+t(\beta)t(\beta^{-1})-g(\beta)g(\beta^{-1})]E^{2}+\mu^{4}-2\mu^{2}[t(\beta)t(\beta^{-1})+g(\beta)g(\beta^{-1})][t(\beta)^{2}-g(\beta)^{2}]t(\beta^{-1})^{2}-g(\beta^{-1})^{2}]=0.
	\label{eq:SM_Poly1}
\end{equation}
For a given $E$, it has two paired solutions $\left(\beta_{1},\beta_{2},\beta_{1}^{-1},\beta_{2}^{-1}\right)$ due to the symmetry of Eq.\,(\ref{eq:SM_Poly1}). The left hand side of Eq.\,(\ref{eq:SM_Poly1}) is an irreducible polynomial in terms of $\beta$ as long as $\mu\neq0$. In other words, this perturbation couples Eqs.\,(\ref{eq:SM_Sol_Poly1}) and (\ref{eq:SM_Sol_Poly2}), which breaks the time-reversal symmetry (\ref{eq:SM_TRS}) together with the Kramers pair. Thus, the presence of the local perturbation breaks the relation of Eq.\,(\ref{eq:GBZ1}). To be precise, we assume $\left|\beta_{1}\right|\leq\left|\beta_{2}\right|\leq1\leq\left|\beta_{2}^{-1}\right|\leq\left|\beta_{1}^{-1}\right|$ without loss of generality, and obtain 
\begin{equation}
	\left|\beta_{2}\right|=\left|\beta_{2}^{-1}\right|\equiv1,
\end{equation}
by applying the standard generalized Brillouin zone. This identity of $|\beta_2|$ implies the breakdown of the skin effect. Therefore, the spectrum dramatically changes in the continuum limit $L\rightarrow\infty$ even if the perturbation is infinitesimal, since the skin effect is a phenomenon of point-gap topology. 

\begin{figure}
	\begin{centering}
		\includegraphics[width=13cm]{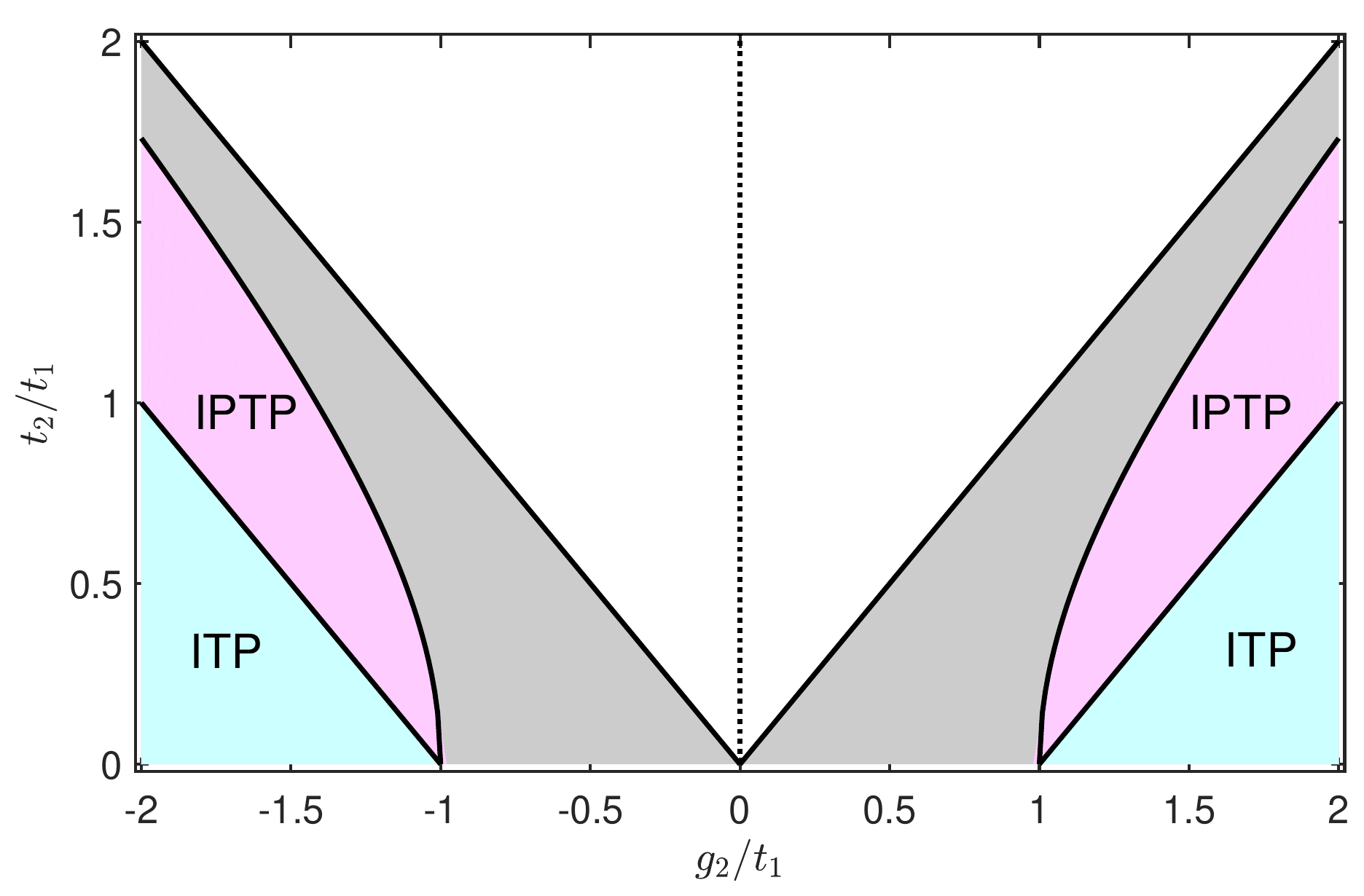}
		\par\end{centering}
	\caption{Phase diagram for the imaginary-gap topology. ITP (cyan) denotes the (conventional) imaginary-gap topological phase while IPTP (purple) is used for the imaginary-gap and point-gap coexisting topological phase. Corresponding to the colorful areas, the system is dynamically unstable under the OBC. 
		\label{fig_S3}}
\end{figure}

\section{Additional discussions on imaginary-gap topology}\label{sec:SM_Imag_Top}

To support the main text, here we provide further discussions on the regime with the imaginary gap opening.  In regime III shown in Figs.\,1(b) and 1(e) of the main text, the large squeezing opens an imaginary gap, and the system's spectrum is formed by two loops located on the imaginary axis in the PBC. The system parameters obey $\Delta^2<-2|t_1t_2-g_1g_2|$ and $|g_1|<|g_2|$. The imaginary-gap topology is characterized by Eq.\,(\ref{W_imaginary}). For $g_1=0$, the condition of opening the imaginary gap is $|g_2|>t_1+t_2$ and the topological phase is always nontrivial, i.e., $W^{\rm (imaginary)}=1$. Then the imaginary-gap topological phase (ITP) corresponds to the cyan areas shown in Fig.\,\ref{fig_S3} and dark gray areas in Fig.\,3 of the main text. In this phase, there are two pairs of zero modes when the OBC is presented. In the basis of canonical coordinates and momenta, these zero modes are ($\hat{x}_{\rm L}^{-s},\hat{x}_{\rm R}^{-s}$) and ($\hat{p}_{\rm L}^{s},\hat{p}_{\rm R}^{s}$) given by Eq.\,(\ref{eq:SM_Sqz_ZM2}). 

Interestingly, in the case where the point gap opens at zero energy and the $\mathbb{Z}_2$ index $\nu(E=0)=1$ is nontrivial, a pair of zero modes ($\hat{x}_{\rm L}^{+},\hat{x}_{\rm R}^{+}$) or ($\hat{p}_{\rm L}^{+},\hat{p}_{\rm R}^{+}$) may survive in the OBC while the imaginary gap is closed in the PBC. It implies that $W^{(\rm imaginary)}$ is unable to {\it fully} predict the occurrence of zero modes, indicating the failure of the bulk-boundary correspondence. To reestablish this correspondence, one can utilize the generalized Brillouin zone performed in Sec.\,\ref{sec:SM_NonBloch}. The continuum bands for this case is given by Eq.\,(\ref{eq:SM_Deform}). It is shown that, compared with the closed loops in the PBC, this spectrum changes from closed loops to open curves, which is consistent with the case in the OBC. In particular, for $g_1=0$, the condition of opening an imaginary gap is altered from $|g_2|>t_1+t_2$ to $|g_2|>\sqrt{t_1^2+t_2^2}$. With respect to this gap, the phase diagram of the imaginary-gap topology is determined by the reconstructed winding number 
\begin{equation}
	\tilde{W}^{\rm (imaginary)}=\frac{1}{2\pi i}\int_{{\rm BZ}}\left(\tilde{p}(k)\right)^{-1}{\rm d}\tilde{p}(k)
	=\begin{cases}
		1, & t_1<\tilde{g}_2\\
		0, & t_1>\tilde{g}_2
	\end{cases},
	\label{tildeW_Imaginary}
\end{equation}
where $\tilde{p}(k)=t_1+\tilde{g}_2e^{ik}$. This reconstructed winding number $\tilde{W}^{\rm (imaginary)}$ is fully capable of characterizing the occurrence of the zero modes. It is nonvanishing for $|g_2|>\sqrt{t_1^2+t_2^2}$, which corresponds to both the cyan and purple areas shown in Fig.\,\ref{fig_S3}. Compared with the cyan areas predicted by $W^{(\rm imaginary)}$, the point gap topology is also nontrivial in the purple areas. It indicates the emergence of an imaginary- and point-gap coexisting topological phase denoted by the IPTP. In this phase, the ${\mathbb{Z}}_{2}$ skin effect greatly inhibits the occurrence of a pair of zero modes, either $(\hat{x}_{\rm L}^{-},\hat{x}_{\rm R}^{-})$ for $g_2>0$ or $(\hat{p}_{\rm L}^{-},\hat{p}_{\rm R}^{-})$ for $g_2<0$, due to the localization competition discussed in Sec.\,\ref{subsec:ZM}. 

Note that, the system under the OBC is always in the dynamical-instability regime, when the squeezing strength is larger than the hopping (i.e., $g_{1,2}>t_{1,2}$), corresponding to the colorful areas in Fig.\,\ref{fig_S3}. It implies that both the nontrivial imaginary-gap topological phase and the coexisting phase belong to this regime.

\section{Additional discussion of detecting topology via the power spectral density\label{sec:SM_Power_Spec}}

\begin{figure}
	\begin{centering}
		\includegraphics[width=17cm]{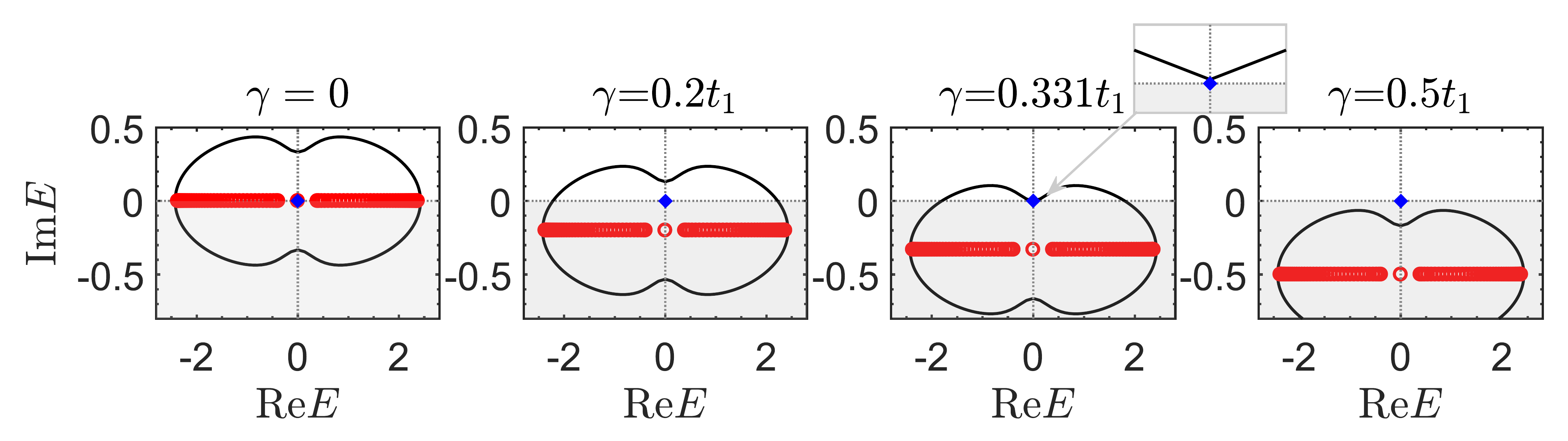} 
		\par\end{centering}
	\caption{Spectrum of the open quantum system as varying $\gamma$. The black curves and red circles represent the spectra in the PBC and OBC, respectively. In the absence of dissipation, the system spectrum under the PBC encloses zero energy and the zero modes occur in the OBC, corresponding the real- and point-gap coexisting topological phase. As increasing $\gamma$, the spectrum moves to the lower half plane and zero energy goes out of the black curve eventually. The criticality happens at $\gamma=\gamma_c\approx 0.331t_1$ as mentioned in the main text. In all plots, $t_2=1.5t_1$, $g_1=0$ and $g_2=0.6t_1$ corresponding to the system in the real- and point-gap coexisting topological phase (the green area of Fig.\,3 of the main text).  
		\label{fig_S4}}
\end{figure}

In this section, let us show the detailed derivation and further discussions regarding the detection of topology via the power spectral density in the presence of dissipation. In the Born-Makrov approximation, the dynamics of system coupled to a reservoir is governed by the Lindblad master equation\,\citep{Gardiner2004} 
\begin{equation}
	\partial_{t}\tilde{\rho}= i\left[\tilde{\rho},\hat{H}_{{\rm SSH}}\right]+\gamma\sum_{j,\sigma}\left(2\hat{a}_{j\sigma}\tilde{\rho}\hat{a}_{j\sigma}^{\dagger}-\left\{ \hat{a}_{j\sigma}^{\dagger}\hat{a}_{j\sigma},\tilde{\rho}\right\} \right), 
	\label{eq:SM_Lindblad_SSH}
\end{equation}
where $\gamma$ is the decay rate at each site and $\tilde{\rho}$ is the density operator of the system. From Eq.\,(\ref{eq:SM_Lindblad_SSH}), the first moments are given by the $4L$-dimensional vector
\[
\hat{\Phi}(t)=\left(\hat{a}_{1A}(t),\hat{a}_{1B}(t),\ldots,\hat{a}_{LB}(t),\hat{a}_{1A}^\dagger(t),\hat{a}_{1B}^\dagger(t),\ldots,\hat{a}_{LB}^\dagger(t)\right)^T,
\]
and second moments are given by the $4L$-by-$4L$ matrix $\hat{M}\left(t\right)=\hat{\Phi}^T\left(t\right)\hat{\Phi}^{\dagger}\left(t\right)$, which satisfies the equations of motion
\begin{align}
	\partial_{t}\hat{\Phi}=&- iH_{{\rm \tau eff}}\hat{\Phi},\label{eq:SM_FM}\\
	\partial_{t}\hat{M}=&- iH_{{\rm \tau eff}}\hat{M}+ i\hat{M}H_{{\rm \tau eff}}^{\dagger}+\gamma\left(\tau^{3}+I\right), 
	\label{eq:SM_SM}
\end{align}
where $H_{\tau{\rm eff}}\coloneqq H_{\tau{\rm SSH}}- i\gamma I$ is the dynamical matrix of the open system. The last term on the right hand side of Eq.\,(\ref{eq:SM_SM}) is originated from the quantum jump. In the steady-state regime, we obtain the vanishing first moments $\Phi_{{\rm  ss}}\coloneqq\lim_{t\rightarrow\infty}\langle \hat{\Phi}\left(t\right)\rangle \equiv0$ and nonvanishing second moments $M_{{\rm ss}}\coloneqq\lim_{t\rightarrow\infty}\langle \hat{M}\left(t\right)\rangle$. The latter is determined by the Lyapunov equation 
\begin{equation}
	- iH_{{\rm \tau eff}}M_{{\rm ss}}+ iM_{{\rm ss}}H_{{\rm \tau eff}}^{\dagger}+\gamma\left(\tau^{3}+I\right)=0.\label{eq:SM_Lyap}
\end{equation}
To obtain the power spectral density of the system at steady state, we consider the time-delayed correlation function $C_{\rho\theta}\left(t,\tau\right)\coloneqq\langle \hat{\rho}\left(t+\tau\right)\hat{\theta}\left(t\right)\rangle $ with the linear observables $\hat{\rho}\left(t\right)=\bm{\rho}^{T}\hat{\Phi}\left(t\right)$ and $\hat{\theta}=\bm{\theta}^{T}\hat{\Phi}\left(t\right)$ ($\bm{\rho}$ and $\bm{\theta}$ are $4L$-dimentional column vectors). From Eqs.\,(\ref{eq:SM_FM}, \ref{eq:SM_SM}), we obtain 
\begin{equation}
	C_{\rho\theta}\left(t,\tau\right)=\bm{\rho}^{\dagger}e^{- iH_{{\rm \tau eff}}\tau}\left\langle M\left(t\right)\right\rangle \bm{\theta}.
\end{equation}
By transforming into the frequency domain and taking $t\rightarrow\infty$, the steady-state normalized power spectral density of $\hat{\rho}$ and $\hat{\theta}$ is obtained
\begin{equation}
	S_{\rho\theta}\left(\omega\right)=\frac{ i\bm{\rho}^{\dagger}G\left(\omega\right)\tau^{3}M_{{\rm ss}}\bm{\theta}}{\lim_{t\rightarrow\infty}C_{\rho\theta}\left(t,0\right)}.
\end{equation}
Here $G\left(\omega\right)\coloneqq\left(\omega I-H_{\tau{\rm eff}}\right)^{-1}\tau^{3}$ is the matrix form of Green's function of the system in the frequency domain, which is defined by 
\begin{equation}
	G\left(i,j;\omega\right)\coloneqq- i\int_{-\infty}^{\infty}dte^{- i\omega t}\Theta\left(t\right)\left\langle \left[\hat{\Phi}_{i}\left(t\right),\hat{\Phi}_{j}^{\dagger}\left(0\right)\right]\right\rangle,
	\label{eq:SM_Green1}
\end{equation}
with $\Theta\left(t\right)$ being the Heaviside step function and $i,j=1,\ldots,4L$. In the input-output formalism, the Green's function indicates the response of systems from an input field with frequency $\omega$. 
\begin{figure}
	\begin{centering}
		\includegraphics[width=16cm]{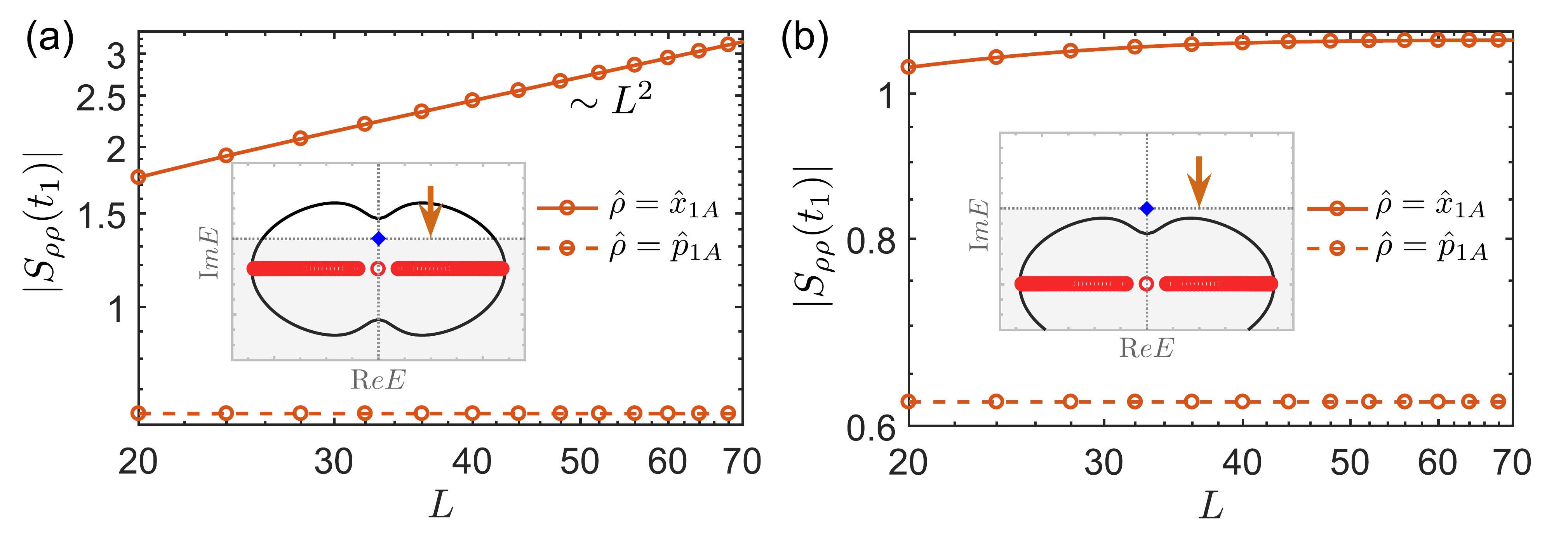} 
		\par\end{centering}
	\caption{The power spectral density $|S_{\rho\rho}(t_1)|$ versus $L$. The decay rate is assumed to be $\gamma=0.2t_1$ in (a) and $\gamma=0.5t_1$(b). The insets plot the system spectra with  associated parameters in the PBC (black curves) and OBC (red circles). The red arrows denote the frequency $\omega=t_1$. Parameters are $t_2=1.5t_1$, $g_1=0$ and $g_2=0.6t_1$ corresponding to the system in the real- and point-gap coexisting topological phase (the green area of Fig.\,3 of the main text).
		\label{fig_S5}}
\end{figure}

To support the claim in the main text on the dissipation-induced topological phase transition, here we focus on the dynamical matrix $H_{\tau\rm eff}$ in Eq.\,(\ref{eq:SM_FM}), which determines the topology of this open quantum system. Under the PBC, we find that the presence of dissipation does not change the time-reversal symmetry (\ref{eq:SM_TRS}). In other words, the dynamical matrix of the open quantum system $H_{\tau\rm eff}(k)$ also has the topological phase like $H_{\tau\rm SSH}(k)$, which is characterized by the $\mathbb{Z}_{2}$ index (\ref{eq:SM_Z2_ind}) by the replacement $H_{{\rm \tau SSH}}\left(k\right)\rightarrow H_{\tau{\rm eff}}\left(k\right)$. It gives us that $\nu\left(E=0\right)=0$ is trivial if $\gamma>\gamma_{c}\equiv\sqrt{g_{2}^{2}-(t_{1}-t_{2})^{2}}$ for $g_1=0$; otherwise, this $\mathbb{Z}_{2}$ index is nontrivial. Physically, as shown in Fig.\,\ref{fig_S4}, increasing the decay rate $\gamma$ moves the spectrum of the squeezed SSH model to the lower plane such that zero energy is out of the loop. Therefore, this process represents a topological phase transition induced by the dissipation. As a result, the skin effect vanishes in the presence of the OBC when $\gamma>\gamma_c$ since the topological invariant is trivial, i.e., $\nu(E=0)=0$. This phase transition corresponds to the disappearance of the dip of $|S_{x_{1A}x_{1A}}(0)|$, which has been shown in Fig.\,4(d) of the main text. 

To see the algebraic divergence of the bulk-state skin effect, as mentioned in the main text, we decompose the Green's function as
\begin{equation}
	\begin{split}G\left(\omega\right)\tau^{3} & =e^{W}\left(\omega I-\tilde{h}_{\tau{\rm SSH}}+ i\gamma I\right)^{-1}e^{-W},\end{split}
	\label{eq:SM_Green2}
\end{equation}
where $\tilde{h}_{\tau{\rm SSH}}\coloneqq h_{{\rm SSH}}\oplus-h_{{\rm SSH}}$ and $e^W=-r_{1}\tau^{1}\sigma^{0}I-r_{2}\tau^{1}\sigma^{3}{\cal L}$ is the squeezing transformation in Eq.\,(\ref{eq:SM_Sqz_SSH}).
The bound of $G\left(\omega\right)\tau^{3}$ is divergent with $L$ for bulk states, which is attributed to $e^{W}$ in Eq.\,(\ref{eq:SM_Green2}). Correspondingly, the bulk-state skin effect manifests itself by the divergence of the power spectral density. Figure\,\ref{fig_S5}(a) shows, as expected, that the peak of $|S_{x_{1A}x_{1A}}(t_1)|$ is algebraically divergent with $L$. For comparison, we increase the decay rate such that the reference frequency $\omega$ goes out the closed curve. In this case, the open system is topologically trivial and the skin effect disappears under the OBC. Therefore, one would expect that the algebraic divergence of $|S_{x_{1A}x_{1A}}(t_1)|$ also breaks down. As shown in Fig.\,\ref{fig_S5}(b), the configuration of $|S_{x_{1A}x_{1A}}(t_1)|$ as varying $L$ verifies our analysis that the divergence arises from the bulk-state skin effect. 

\begin{figure}
	\begin{centering}
		\includegraphics[width=13cm]{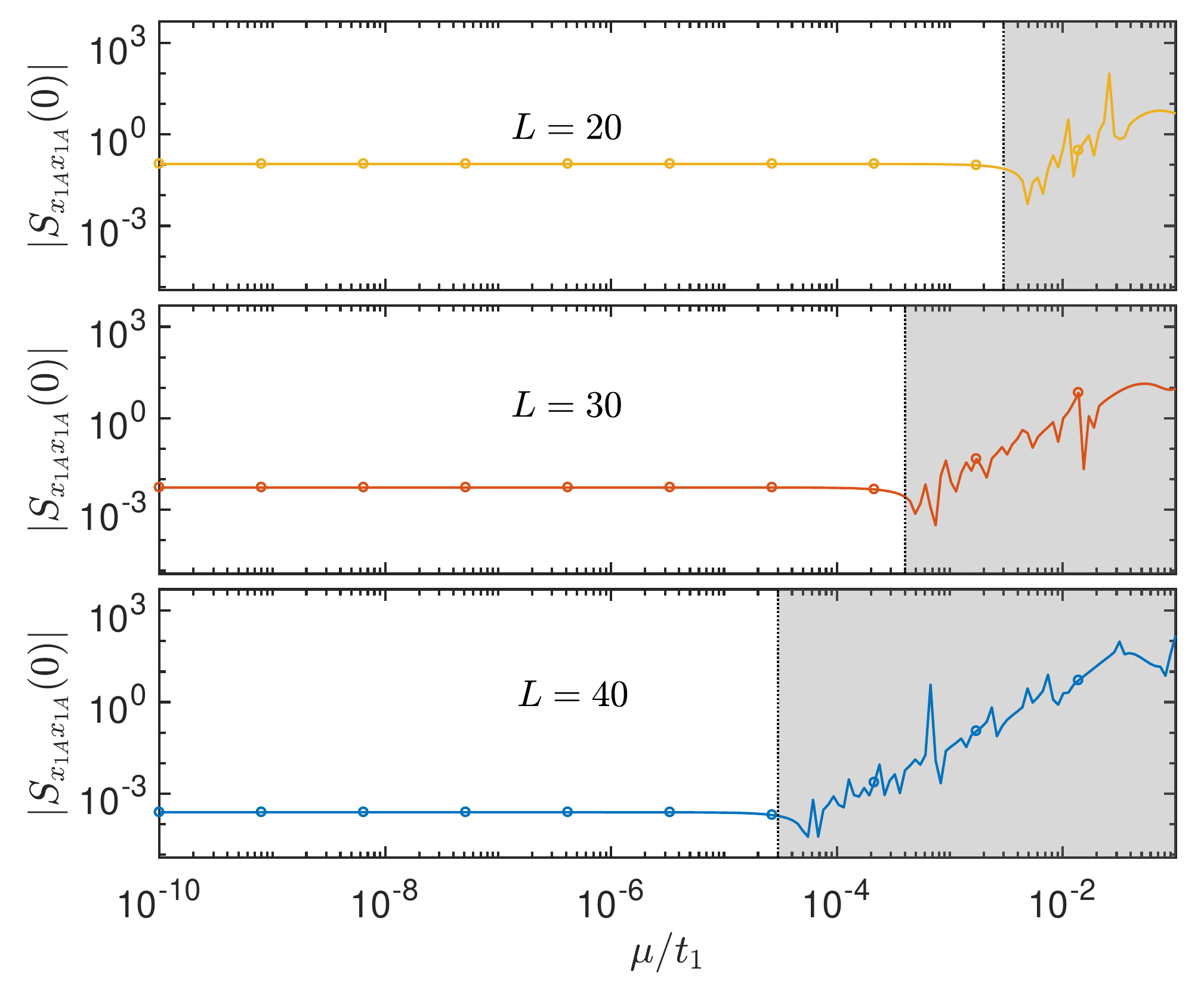}
		\par\end{centering}
	\caption{The power spectral density $\left|S_{x_{1A}x_{1A}}\left(0\right)\right|$ versus the local perturbation strength $\mu$. Here $L=20,30,40$ is set from top to down, respectively, in good agreement with the scaling $\mu/t_1\sim 5\times10^{-3},4\times10^{-4},3\times10^{-5}$. Other parameters are $t_{2}=1.5t_{1}$, $g_{1}=0$, $g_{2}=0.6t_{1}$ and $\gamma=0.2t_{1}$ corresponding to the system in the real- and point-gap coexisting topological phase (the green area of Fig.\,3 of the main text). 
		\label{fig_S6}}
\end{figure}

Moreover, the skin-effect inhibition on zero modes would be destroyed by symmetry-breaking local perturbations. As discussed in the main text, the breakdown of the skin effect can be estimated by the scaling of the perturbation $\mu/t_1\sim\epsilon_2\xi^{-L}$ with $\xi=e^{|r_2|}|\delta|^{1/2}$, which has been given in the main text. This can be identified by the zero-frequency dip of the power spectral density at end sites. Figure\,\ref{fig_S6} plots $|S_{x_{1A}x_{1A}}(0)|$ along with varying $\mu$ for $L=20,30,40$, and shows the instability of this topological inhibition. The disappearance of the inhibition happens at $\mu/t_1\sim 5\times10^{-3},4\times10^{-4},3\times10^{-5}$ for $L=20,30,40$, which is in well agreement with the scaling. 

\section{Correspondence between squeezing transformation and skin effect in QBSs \label{sec:SM_Proof}}
This section is devoted to proving the correspondence between the squeezing transformation and skin effect in any QBSs under the OBC. To be concrete, the skin effect is originated from the nonnormality of the $2N$-by-$2N$ dynamical matrix $H_{\tau}$, i.e., $\left[H_{\tau},H_{\tau}^{\dagger}\right]\neq0$. In other words, the nonormality of $H_{\tau}$ is a necessity (but not sufficient) for the occurrence of the skin effect. Therefore,
we need to prove that any nonormal $H_{\tau}$ can be transformed to a normal matrix by a squeezing transformation, i.e., $K\in{\rm SU}\left(N,N\right)$ obeying $\tau^{1}K^{*}\tau^{1}=K$. This is summarized by the proposition and corollary below. 
\begin{quote}
	\textbf{Proposition I}: \emph{If the QBS is diagonalizable, then there exists some squeezing transformation $K\in{\rm SU}(N,N)$ obeying $\tau^1 K^*\tau^1=K$ such that the dynamical matrix is similar to a normal one, i.e., $K^{-1}H_{\tau}K=O_{\tau}$} with $\left[O_{\tau},O_{\tau}^{\dagger}\right]=0$.
	\label{Prop-I} 
\end{quote}
\emph{Proof}: Let the dynamical matrix $H_{\tau}$ be diagonalizable. We introduce the two sets of right and left eigenstates \citep{Brody2013JPA} in a compact form, i.e., 
\begin{equation}
	H_{\tau}R=RE,\ \ \ H_{\tau}^{\dagger}L=LE^{*},
\end{equation}
where $R=\left(\left|R_{1}\right\rangle ,\ldots,\left|R_{2N}\right\rangle \right)$ and $L=\left(\left|L_{1}\right\rangle ,\ldots,\left|L_{2N}\right\rangle \right)$ form the complete biorthogonal basis $\langle L_{i}\vert R_{j}\rangle=\delta_{ij}$ and $\sum_{j}\left|R_{j}\rangle\langle L_{j}\right|=I$. Thus, it yields the relation $L^{\dagger}R=R^{\dagger}L=LR^{\dagger}=RL^{\dagger}=I$. The non-Hermitian Hamiltonian respects the pseudo-Hermiticity, 
\begin{equation}
	\begin{split}\tau^{3}H_{\tau}\tau^{3} & =H_{\tau}^{\dagger}\\
		\Rightarrow \left(\tau^{3}R\right)E\left(\tau^{3}R\right)^{-1} & =LE^{*}L^{-1}.
	\end{split}
\end{equation}
It implies that $\tau^{3}R$ is another set of $L$, and then it gives
\begin{equation}
	\tau^{3}R=L\chi,
\end{equation}
where $\chi=R^{\dagger}\tau^{3}R$ is unitary and satisfies $\chi^{-1}=\chi^{\dagger}=\chi$ due to the biorthogonality. Then we obtain 
\begin{equation}
	\left(RR^{\dagger}\right)\tau^{3}\left(RR^{\dagger}\right)=\tau^{3}.
\end{equation}
By introducing the polar decomposition $R=KU$ ($K=\sqrt{RR^{\dagger}}$ is positive definite and $U=K^{-1}R$ is unitary), we obtain $K\tau^{3}K=\tau^{3}$ and 
\begin{equation}
	K^{-1}H_{\tau}K=O_{\tau},
\end{equation}
where $O_{\tau}\coloneqq UEU^{\dagger}$ is the normal matrix. $\blacksquare$

Proposition I shows that any nonnormal $H_\tau$ can be transformed to a normal one by a squeezing transformation. It implies that the squeezing transformation $K$ corresponds to the skin effect in the QBS when it exists. Note that the referred QBS in Proposition I can be dynamically stable or unstable. With the virtue of Proposition I, a corollary naturally arises for the dynamically stable QBS in the following. 
\begin{quote}
	\textbf{Corollary II}: \emph{The QBS is dynamically stable if and only if there exists a matrix $V$ such that the associated Hamiltonian $H$ can be pseudounitarily diagonalized, i.e., $V^{\dagger}HV=E$ with $V^{\dagger}\tau^{3}V=V\tau^{3}V^{\dagger}=\tau^{3}$. \label{Corollary-II:}} 
\end{quote}
\emph{Proof}: (i) Assume that the QBS is dynamically stable, that is, the spectrum of the QBS is real, $E\in\mathbb{R}$. In fact, it gives the pseudounitary relation ($V\equiv R$), $V^{\dagger}\tau^{3}V=\tau^{3}$. Then it arrives at $V^{\dagger}HV=E$. (ii) The proof on the inverse proposition is trivial. $\blacksquare$

\emph{Another parallel proof}: (i) Let $H$ be pseudounitarily diagonalizable. It means that there must exist a nonsingular matrix $V$ such that $V^{\dagger}HV=E$, where $V$ obeys the relation $V^{\dagger}\tau^{3}V=\tau^{3}$. And utilizing the Hermiticity yields real spectrum, $E\in\mathbb{R}$. Thus, the QBS is dynamically stable. 

(ii) Let the QBS be dynamically stable. Then there exists a nonsingular matrix $V_{1}$ such that $V_{1}^{-1}H_{\tau}V_{1}=\Lambda$, where the diagonal matrix $\Lambda$ satisfies ${\rm Im}\Lambda=0$. Utilizing the polar decomposition $V_{1}=KU$ with unitary $U=(V_{1}V_{1}^{\dagger})^{-1/2}V_{1}$ and positive definite  $K=\sqrt{V_{1}V_{1}^{\dagger}}$, we have $K^{-1}H_{\tau}K=h$, where $h=UEU^{\dagger}$ is Hermitian. Due to the Hemiticity of $H$, we obtain 
\begin{equation}
	\tau^{3}HK^{2}=K^{2}H\tau^{3}.
	\label{eq:SM_Hermiticity}
\end{equation}
According to Williamson's theorem\,\citep{Williamson1936AJM,Simon1999JMP}, the positive Hamiltonian $H$ can be pseudounitarily diagonalized, i.e., $K^{2}=V_{2}\Pi V_{2}^{\dagger}$, where the matrix $V_{2}$ satisfies the relation $V_{2}\tau^{3}V_{2}^{\dagger}=V_{2}^{\dagger}\tau^{3}V_{2}=\tau^{3}$ and the diagonal $\Pi>0$. Then, substituting $K^{2}= V_{2}\Pi V_{2}^{\dagger}$ back into Eq.\,(\ref{eq:SM_Hermiticity}), we obtain $\tau^{3}(V_{2}^{\dagger}HV_{2})\Pi=\Pi (V_{2}^{\dagger}HV_{2})\tau^{3}$, whose elements obey 
\begin{equation}
	\left(V_{2}^{\dagger}HV_{2}\right)_{mn}\left(\tau_{m}^{3}\Pi_{n}-\tau_{n}^{3}\Pi_{m}\right)=0.
\end{equation}
Note that $\tau^{3}=I\oplus-I$. If $m=1,\ldots,N$ and $n=N+1,\ldots,2N$, thus, $\left(\tau_{m}^{3}\Pi_{n}-\tau_{n}^{3}\Pi_{m}\right)<0$, requiring $\left(V_{2}^{\dagger}HV_{2}\right)_{mn}=0$. Then, it reveals that 
\begin{equation}
	V_{2}^{\dagger}HV_{2}=\left(\begin{array}{cc}
		h_{1} & 0\\
		0 & h_{2}
	\end{array}\right)=\left(\begin{array}{cc}
		U_{1}E_{1}U_{1}^{\dagger} & 0\\
		0 & U_{2}E_{2}U_{2}^{\dagger}
	\end{array}\right).
\end{equation}
Finally, we arrive at the pseudounitary diagonalization of $H$, 
\begin{equation}
	V^{\dagger}HV=\left(\begin{array}{cc}
		E_{1} & 0\\
		0 & E_{2}
	\end{array}\right),
\end{equation}
where $V\coloneqq V_{2}\left(\begin{array}{cc}
	U_{1} & 0\\ 
	0 & U_{2}
\end{array}\right)$ obeys $V^{\dagger}\tau^{3}V=V\tau^{3}V^{\dagger}=\tau^{3}$. $\blacksquare$

In the above, we have proved that the nonnormal matrix $H_\tau$ can be transformed to a normal one by some squeezing transformation. In particular, the obtained normal matrix is block-diagonal for any dynamically stable QBSs in Corollary II. We naturally concludes that the skin effect for any QBSs corresponds to a squeezing transformation.  

\section{Additional discussion on intercell squeezing}\label{sec:SM_ISqz}

In this section, we provide additional discussions on the intercell squeezing mentioned in the main text. It is referred to the bi-particle annihilation and/or creation process between intercell modes, i.e., $\sum_{j}g_2(\hat{a}_{jB}\hat{a}_{j+1A}+{\rm H.c.})$. It describes a nondegenerate parametric amplification, and can introduce the entanglement between two bosonic modes in the adjacent cells. 

Without loss of generality, here we consider a two-mode Hamiltonian $\hat{H}=g_2(\hat{a}_{jB}\hat{a}_{j+1A}+{\rm H.c.})$ to illustrate the mechanism of the squeezing, due to the fact that the intercell squeezing is spatially separable and identical in the proposed 1D lattice system. Suppose the initial state of the system is a vacuum state, $\vert\psi(0)\rangle = \vert 00\rangle$. At time $t$, the system is described by a two-mode squeezed state,  $\vert\psi(t)\rangle =e^{-ig_2t (\hat{a}_{jB} \hat{a}_{j+1A} + {\rm H.c.})}\vert 00\rangle$. Note that $g_2$ determines the squeezing strength of the state. Such a state can be expressed in the Fock space as\,\citep{Lvovsky2015}
\begin{equation}
	\left\vert\psi(t)\right\rangle=\sum_{n=0}^{\infty}\frac{(-i\tanh g_2t)^{n}}{\cosh g_2t}\vert nn\rangle,
	\label{eq:SM_TMSqz}
\end{equation}
where $\vert nn\rangle$ is tensor product of the Fock state $\vert n\rangle$. It can be seen that $\vert\psi(t)\rangle$ is an entangled state. To quantify the entanglement, here we introduce the von Neumann entropy, by definition, $S\coloneqq -{\rm Tr}(\tilde{\rho}_a\log_2\tilde{\rho}_a)$, where $\tilde{\rho}_a\coloneqq {\rm Tr}_{b}(\tilde{\rho})$ is the reduced density matrix of mode $\hat{a}_{j+1A}$ and ${\rm Tr}_b$ is the partial trace over mode $\hat{a}_{jB}$. Here $\tilde{\rho}$ is the density matrix of the two-mode system. The entropy is obtained by 
\begin{equation}
	S=\sum_{n=0}^{\infty}\frac{\tanh^{2n} g_2t}{\cosh^2 g_2t}\log_2{\frac{\tanh^{2n} g_2t}{\cosh^2 g_2t}}.
	\label{eq:SM_Entropy}
\end{equation}
Alternatively, this entanglement can also be quantified by the logarithmic negativity, by definition, $E_N\coloneqq \log_2{\Vert\tilde{\rho}^{\Gamma_a}}\Vert_{\rm T}$, where $\Gamma_a$ denotes the partial transpose on mode $\hat{a}_{j+1A}$ and $\Vert\cdot\Vert_{\rm T}$ is the trace norm. After some calculation, we obtain $E_N=2g_2t\log_2 e$. The nonzero entropy $S$ and logarithmic negativity $E_N$ for any time $t>0$ clearly indicates that the entanglement between the two modes $\hat{a}_{j+1A}$ and $\hat{a}_{jB}$ is introduced into our lattice system, which originally comes from the intercell squeezing. 

\begin{figure}
	\begin{centering}
		\includegraphics[width=18cm]{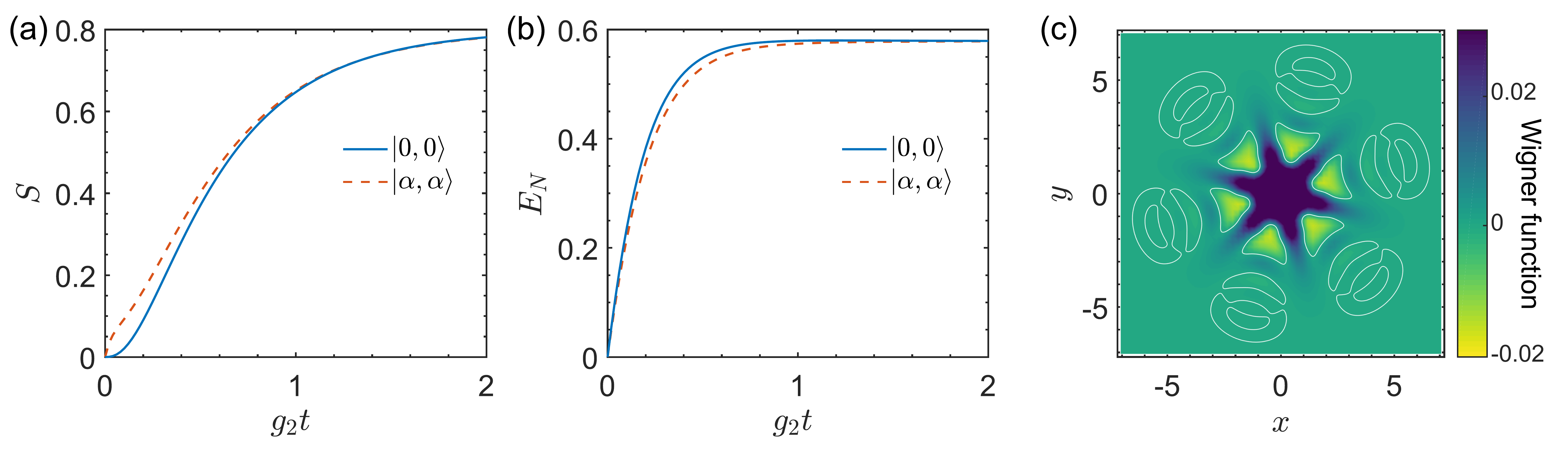}
		\par\end{centering}
	\caption{ (a) von Neumann entropy $S$ and (b) logarithmic negativity $E_N$ versus time $t$. The initial state is prepared as the vacuum state (blue solid curves) and coherent state $\vert\alpha,\alpha\rangle$ with $\alpha=1$ (red dashed curves). (c) Wigner function of the intercell squeezing model at steady state. The contour curves (white) indicate the zero value of the steady-state Wigner function. The dissipation rate is $\gamma=2g_{2}$. 
		\label{fig_S7}}
\end{figure}

So far, the case considered is ideal. For practical situations, one should consider the influence of dissipation to the system. In Figs.\,\ref{fig_S7}(a) and \ref{fig_S7}(b), we plot the entropy $S$ and logarithmic negativity $E_N$ for the two-mode system by calculating the master equation $\partial_t\tilde{\rho}=-i[\hat{H},\tilde{\rho}]+\gamma\sum_{l=jB, j+1A}(2\hat{a}_{l}\tilde{\rho}\hat{a}^{\dagger}_{l}-\{\hat{a}_{l}^\dagger\hat{a}_{l},\tilde{\rho}\})$ with the decay rate $\gamma$. These quantities approach to nonzero values as varying time even in presence of strong dissipation, i.e., $g_2<\gamma$. In the steady-state regime, $S$ and $E_N$ is still nonzero. Moreover, this nonclassicality of the steady state can be identified by the Wigner function. In Fig.\,\ref{fig_S7}(c), we plot the steady-state Wigner function and its negative value indicates faithfully the nonclassicality of the steady state. 

In addition, the intercell squeezing can be typically obtained in the real system by employing the three-wave mixing interaction, which is discussed illustratively in Sec.\,\ref{sec:SM_Implementation}. 

\section{Physical implementation of the squeezed SSH model}\label{sec:SM_Implementation}

This section is devoted to in detail discussing a physical implementation of the QBS with Hamiltonian (3) in the main text. 

\begin{figure}
	\begin{centering}
		\includegraphics[width=14cm]{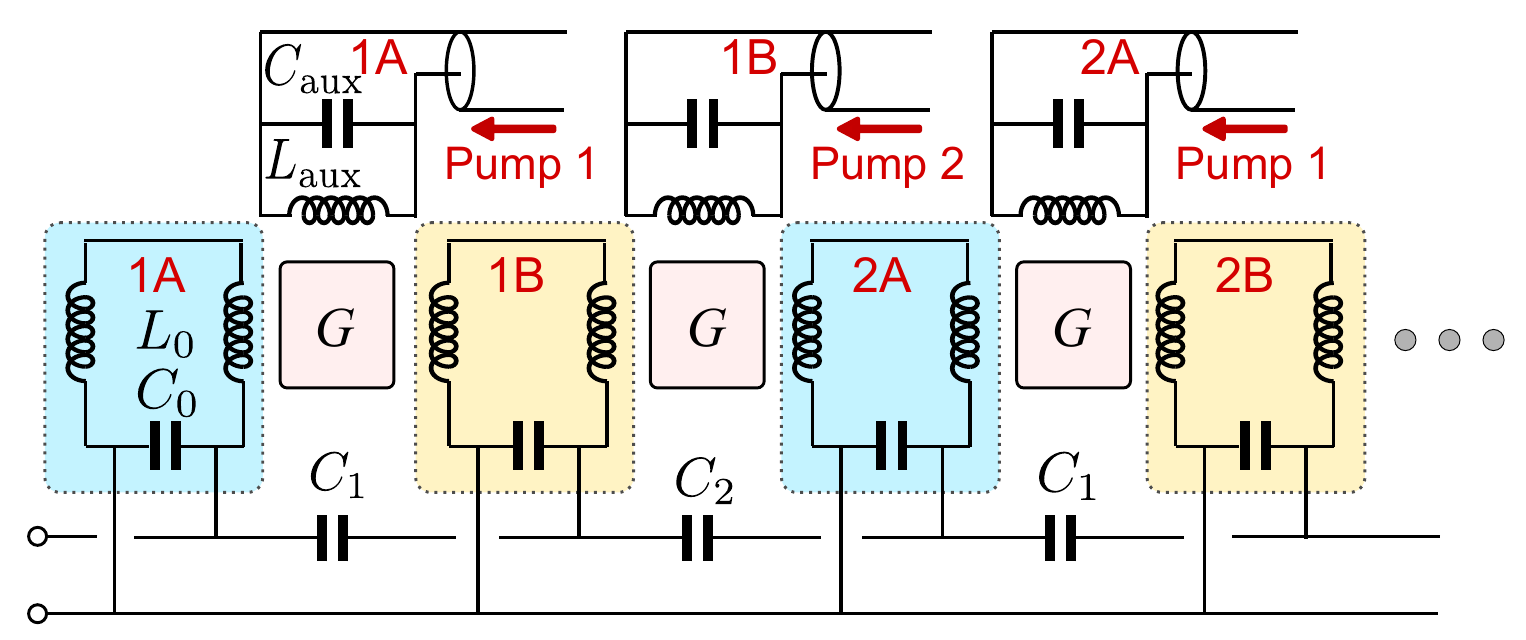}
		\par\end{centering}
	\caption{  Schematic of physically implementing the proposed 1D lattice [see Fig. 1(a) in the main text]. Blue (yellow) boxes depict the L LC oscillators hosting local modes with annihilation operator $\hat{a}_{jA}$ ($\hat{a}_{jB}$) and frequency $\omega_{a}=\sqrt{1/2L_0C_0}\sqrt{(C_0+C_1+C_2)/C_0}$. They form a dimer chain and the hopping rates between the intracell and intercell sites are given by $t_{1,2}=\sqrt{1/2L_0C_0}\sqrt{C_{1,2}/C_0}$, respectively. In particular, the crucial intracell and intercell squeezings can be implemented via the three-wave mixing interaction. The red boxes $G$ denoting the Josephson ring modulator or SNAIL device is coupled to the adjacent LC circuits by mutual inductances and introduce the nondegenerate three-wave mixing interaction with form $g\hat{\Phi}_X\hat{\Phi}_Y\hat{\Phi}_Z$. The $(2L-1)$ auxiliary LC oscillators hosting the bosonic modes with annihilation operator $\hat{b}_{j\sigma}$ ($\sigma=A,B$) and frequency $\omega_{\rm aux}=\sqrt{1/L_{\rm aux}C_{\rm aux}}$ are pumped coherently.
		\label{fig_S8}}
\end{figure}

\subsection{Implementation based on the circuit QED }

As shown in Fig.\,\ref{fig_S8}, $2L$ LC oscillators with inductance $2L_0$ and capacitance $C_0$ is alternately coupled to capacitors $C_1$ and $C_2$, forming a dimer chain (SSH model). The unit cell of the chain consists of two sites A (blue) and B (yellow). The charge on the capacitor of a LC oscillator at $j\sigma$ can be quantized introducing bosonic annihilation and creation operators, $\hat{Q}\propto(\hat{a}_{j\sigma}$ + $\hat{a}^\dagger_{j\sigma})$. Here $j$ denotes the unit cell and $\sigma=A,B$ labels the sublattice. The capacitive couplings between the adjacent LC oscillators are $\propto\hat{Q}_{jA}\hat{Q}_{jB}$ and $\propto\hat{Q}_{jB}\hat{Q}_{j+1A}$. 

In particular, the crucial intracell and intercell squeezing can be implemented in the superconducting circuits via the three-wave mixing interaction. Beside the dimer chain, we introduce $(2L-1)$ auxiliary LC oscillators ($L_{\rm aux}, C_{\rm aux}$) hosting local bosonic modes $\hat{b}_{j\sigma}$  with frequency $\omega_{\rm aux}=2\omega_{a}$, to generate the three-wave mixing interactions. Now the total Hamiltonian of system are $\hat{H}_{\rm tot}=\hat{H}_0+\hat{H}_{\rm hop}+\hat{H}_{\rm int}$, where $\hat{H}_{0}=\sum_{j\sigma}\omega_{a}\hat{a}_{j\sigma}^\dagger\hat{a}_{j\sigma}+\omega_{\rm aux}\hat{b}_{j\sigma}^\dagger\hat{b}_{j\sigma}$ is the free LC oscillators. $\hat{H}_{\rm hop}$ describes the hopping of the dimer chain of LC oscillators and can be expressed as 
\begin{equation}
	\hat{H}_{\rm hop} =\sum_{j=1}^{L}\left[t_1(\hat{a}_{jA}+\hat{a}_{jA}^\dagger)(\hat{a}_{jB}+\hat{a}_{jB}^\dagger)+t_2({{\hat{a}}_{jB}+\hat{a}}_{jB}^\dagger)({\hat{a}}_{j+1A}+{\hat{a}}_{j+1A}^\dagger)\right],   \label{eq:SM_PI_1}
\end{equation}
where the hopping rates are given by $t_{1,2}=\sqrt{1/2L_0C_0}\sqrt{C_{1,2}/C_0}$. Here $\hat{H}_{\rm int}$ describes the nondegenerate three-wave mixing interaction and can be introduced by the Josephson ring modulator\,\citep{Abdo2013PRB} or the Superconducting Nonlinear Asymmetric Inductive eLement (SNAIL)\,\citep{Frattini2017APL}, represented as the red boxes $G$ in Fig.\,\ref{fig_S8}. Specifically, mutual inductances couple the $G$ circuit to the three adjacent inductances (two in the dimer chain and one in the auxiliary LC oscillators) via the inductances of the circuit, which leads to the nondegenerate three-wave mixing interaction with form $g\hat{\Phi}_X\hat{\Phi}_Y\hat{\Phi}_Z$. Here $\hat{\Phi}_{X,Y,Z}$ denote the flux of three adjacent LC oscillators, and the coupling strength $g$ is determined by $L_0$, $L_{\rm aux}$ and the mutual inductance between them. Then the nonlinear interaction between two LC oscillators in the dimer chain and one adjacent auxiliary LC oscillator can be expressed as 
\begin{equation}
	\hat{H}_{\rm int} = \sum_{j=1}^{L}\left[ig_0(\hat{a}_{jA}-\hat{a}_{jA}^\dagger)(\hat{a}_{jB}-\hat{a}_{jB}^\dagger)(\hat{b}_{jA}-\hat{b}_{jA}^\dagger)+ig_0(\hat{a}_{jB}-\hat{a}_{jB}^\dagger)(\hat{a}_{j+1 A}-\hat{a}_{j+1 A}^\dagger)(\hat{b}_{jB}-\hat{b}_{jB}^\dagger)\right], \label{eq:SM_PI_2}
\end{equation}
where $g_0=g\sqrt{XYZ/8}$ with $X=Y=\sqrt{2L_0/C_0}$ and $Z=\sqrt{L_{\rm aux}/C_{\rm aux}}$. 
Under the rotating wave approximation, Eq.\,(\ref{eq:SM_PI_1}) is reduced to the conventional SSH model, i.e., 
\begin{equation}
	\hat{H}_{\rm hop}'=\sum_{j=1}^{L}\left(t_1\hat{a}_{jA}^\dagger\hat{a}_{jB}+t_2\hat{a}_{jB}^\dagger{\hat{a}}_{j+1A}+{\rm H.c.}\right). 
\end{equation}
Similarly, the nonlinear interaction (\ref{eq:SM_PI_2}) is then reduced to 
\begin{equation}
	\hat{H}_{\rm int}'=\sum_{j=1}^{L}ig_0\left(\hat{a}_{jA}\hat{a}_{jB}\hat{b}_{jA}^\dagger +\hat{a}_{jB}\hat{a}_{j+1A}\hat{b}_{jB}^\dagger+{\rm H.c.}\right). 
	\label{eq:SM_PI_3}
\end{equation}
To meet our goal, we pump each auxiliary cavity coherently with $\langle\hat{b}_{j\sigma}\rangle = i\beta_{\sigma} e^{-i\omega_{\rm aux}t}$ ($\beta_\sigma\in \mathbb{R}$), and the nonlinear interaction  can be further reduced as, 
\begin{equation}
	\hat{H}_{\rm int}''=\sum_{j=1}^{L}\left(g_1\hat{a}_{jA}\hat{a}_{jB}e^{i\omega_{\rm aux}t}+g_2\hat{a}_{jB}\hat{a}_{j+1A}e^{i\omega_{\rm aux}t}+{\rm H.c.}\right), 
	\label{eq:SM_PI_4}
\end{equation}
where $g_1=g_0\beta_{A}$ and $g_2=g_0\beta_{B}$. In a rotating frame at frequency $\omega_a$, Eq.\,(\ref{eq:SM_PI_4}) becomes time-independent, and the effective Hamiltonian $\hat{H}_{\rm eff}=(\hat{H}_{\rm hop}'+\hat{H}_{\rm int}'')$ shares the same formula of Eq.\,(3) in the main text. 

\subsection{Accessible parameter regime}
Our proposal relies on the three-wave mixing interaction. It can be directly implemented in the microwave regime of superconducting quantum circuits like the Josephson ring modulator\,\citep{Abdo2013PRB} and SNAIL device\,\citep{Frattini2017APL}. As discussed in the above, those superconducting-circuit elements can introduce the mutual inductance between the dimer chain and auxiliary LC oscillators, which results in the three-wave mixing process eventually.  

The required energy hierarchy for our proposal is 
\begin{equation}
	\gamma<g_{1},g_{2}< t_{1},t_{2},
	\label{eq:SM_PI_6}
\end{equation}
where $\gamma$ is the decay rate of the cavity mode $\hat{a}_{j\sigma}$.
In general, the nonlinear interaction strength $g_0$ is much smaller than the decay rate $\gamma$. The relevant parameters satisfy the relation
\begin{equation}
	g_0\ll \gamma< t_{1},t_{2}\ll\omega_{a}<\omega_{\rm aux}=2\omega_{a}.
	\label{eq:SM_PI_6}
\end{equation}
For the Josephson ring modulator\,\citep{Abdo2013PRB}, the system parameters are given by $\omega_a\sim 2\pi\times 10$GHz, $Q\sim 100$, $g_0\sim 2\pi\times 1$MHz, where $Q$ denotes the quality factor of the cavity modes $\hat{a}_{j\sigma}$. The decay rate $\gamma$ is $\sim 2\pi\times 100$MHz. Since the photon number of the pumped auxiliary modes can reach to the order of $\langle \hat{b}_{j\sigma}^\dagger\hat{b}_{j\sigma}\rangle \sim10^4$, the squeezing strength $g_{1,2}$ is comparable with $t_{1,2}$ and $\gamma$, which brings us to the desired parameter regime $\gamma<g_{1},g_{2}< t_{1},t_{2}$. 


\begin{thebibliography}{102}
\bibitem [{Con()}]{Continuum}%
	\BibitemOpen
	\href@noop {} {}\bibinfo {note} {Here the continuum limit refers to the fact
		that both ends of the chain go to infinity, which is the so-called infinite
		boundary condition. One should not be confused with the semi-infinite
		boundary condition~\citep{Gong2018PhysRevX, Okuma2020PRL}. Moreover, the
		infinite boundary condition is equivalent to the PBC by taking $L\rightarrow
		\infty$~\citep{Trefethen2005}.}\BibitemShut {Stop}%
\end{thebibliography}
\end{document}